\documentclass[11pt,a4paper]{article}
\pdfoutput=1
\usepackage{jheppub}
\usepackage{amsmath,amsfonts,amssymb,bbm, mathtools, mathrsfs}
\usepackage{hyperref, csquotes}
\usepackage{graphicx, color}
\usepackage{tikz,pgf,tikz-cd,float}
\usepackage{physics,tensor}
\include{fdiagrams}

\newcommand{\one}{\mathbbm 1}
\newcommand{\R}{\mathbb R}
\newcommand{\C}{\mathbb C}
\newcommand{\e}{\textrm{e}}

\newcommand{\cas}{\textrm{Cas}}     
\newcommand{\SU}{\text{SU$(2)$}}
\newcommand{\SL}{\text{SL$(2,\C)$}}

\newcommand{\su}{\mathfrak{su}(2)}
\newcommand{\spl}{\mathfrak{sl}\left(2,\C\right)}
\newcommand{\HH}{\text{H}^3}
\newcommand{\defeq}{\vcentcolon=}
\newcommand{\eqdef}{=\vcentcolon}

\renewcommand{\hom}{\mathfrak{hom}(2)}
\newcommand{\Hom}{\text{Hom}(2)}

\begin{document}

\title{Emergent Cosmology from Quantum Gravity in the Lorentzian Barrett-Crane Tensorial Group Field Theory Model}

\author{Alexander F. Jercher,}
\author{Daniele Oriti,}
\author{Andreas G. A. Pithis}
\emailAdd{alexander.jercher@campus.lmu.de, daniele.oriti@physik.lmu.de, andreas.pithis@physik.lmu.de}

\affiliation{Arnold Sommerfeld Center for Theoretical Physics,\\ Ludwig-Maximilians-Universit\"at München \\ Theresienstrasse 37, 80333 M\"unchen, Germany, EU}
\date{\today}

\begin{abstract}
{
We study the cosmological sector of the Lorentzian Barrett-Crane (BC) model coupled to a free massless scalar field in its Group Field Theory (GFT) formulation, corresponding to the mean-field hydrodynamics obtained from coherent condensate states. The relational evolution of the condensate with respect to the scalar field yields effective dynamics of homogeneous and isotropic cosmologies, similar to those previously obtained in $\text{SU}(2)$-based EPRL-like models. Also in this manifestly Lorentzian setting, in which only continuous $\text{SL}(2,\mathbb{C})$-representations are used, we obtain generalized Friedmann equations that generically exhibit a quantum bounce, and can reproduce all of the features of the cosmological dynamics of EPRL-like models. This lends support to the expectation that the EPRL-like and BC models may lie in the same continuum universality class, and that the quantum gravity mechanism producing effective bouncing scenarios may not depend directly on the discretization of geometric observables.
\newline
}
\end{abstract}

\maketitle

\section{Introduction}\label{sec:Introduction}

Theories of quantum gravity proposing new fundamental building blocks of quantum spacetime, and not constructed directly in terms of continuum fields, are confronted with the major challenge of extracting macroscopic physics in a proper continuum limit that agrees (approximately) with general relativity (GR). This is not only a necessary consistency check and an important connection to existing observations, but also a first step for obtaining then new falsifiable predictions about deviations from classical GR and other quantum gravity-induced phenomena. 

Cosmology, understood as the physics of homogeneous and close-to homogeneous universes, represents an appropriate and simple enough setting for studying quantum gravity effects in a macroscopic approximation. First, spatial homogeneity yields a drastic simplification of the configuration space resulting in feasible computations. Second, quantum fluctuations are expected to dominate at early times of the universe, possibly leading to observable signatures e.g. in the cosmic microwave background~\cite{Ashtekar:2021kfp} or in gravitational waves~\cite{Calcagni:2019kzo}. Much of modern quantum gravity phenomenology has to do with quantum gravity effects that could also affect cosmological physics~\cite{Addazi:2021xuf} 

Starting from the fundamental quantum geometrical degrees of freedom, two possibilities to resulting in a theory with only those corresponding to homogeneous data at macroscopic scales present themselves\footnote{In fact, the same alternatives are present in classical GR~\cite{Buchert:2007ik,Buchert:2015iva}}: symmetry reduction and coarse-graining. Symmetry reduction refers to reducing the degrees of freedom on the kinematical level such that the configuration space contains only the relevant macroscopic observables, with consequent (exact) reduction of the dynamics to these degrees of freedom only. This is the route chosen in quantum cosmology~\cite{Ashtekar:2011ni}, where the symmetry reduction is performed at the classical level, and the resulting reduced data are then quantized. It is also the strategy adopted in some recent approaches trying to connect fundamental loop quantum gravity to cosmology, starting instead from the quantum states state space of the full theory~\cite{Alesci:2016gub}. By coarse-graining, many microscopic degrees of freedom are collectively described by a few macroscopic parameters, in correspondence with relevant macroscopic observables, but with the effective dynamics for them being (possibly very much) affected by the presence of the entirety of dynamical degrees of freedom. An example of a straightforward result of coarse-graining is the mean-field approximation in condensed matter physics, which provides also a template for our strategy for connecting quantum gravity to cosmology. In fact, symmetry reduction is a special case of extremely drastic coarse-graining, where all degrees of freedom not corresponding directly (e.g. as eigenstates) to the relevant macroscopic observables are completely ignored, rather than approximately dealt with or resummed into effective ones~\cite{Oriti:2016qtz}. 

A particular compelling framework in which the idea of extracting cosmology from a coarse-graining of the full quantum gravity dynamics is the Group Field Theory (GFT) formalism~\cite{Freidel:2005qe,Oriti:2011jm,Oriti:2007qd}. GFTs are quantum and statistical field theories defined on a Lie group, with the characterizing property of non-local field interactions. One of the many perspectives that can be taken on GFTs is that they are a generalization of matrix models~\cite{DiFrancesco:1993cyw,David:1992jw} for $2d$ quantum gravity. From this point of view, they are examples of the more general {\it tensorial group field theory formalism} (TGFT), including all tensorial field theories sharing the same non-local type of interactions, starting form the simplest random tensor models~\cite{Gurau:2011xp,GurauBook}, i.e. the immediate combinatorial generalization of matrix models. From this point of view, GFTs represent the specific subset of TGFT models characterized by additional \enquote{quantum geometric data} providing quantum states and dynamics with enriched discrete geometric content, facilitating their geometric and physical interpretation (at the cost of complicating their mathematical treatment). Also, thanks to these additional data, GFTs provide a link between several quantum gravity approaches, including canonical loop quantum gravity (LQG)~\cite{Thiemann:2007pyv,Rovelli:2004tv,Ashtekar:2004eh,Ashtekar:2021kfp}, spin foam models~\cite{Perez:2003vx,Perez:2012wv,Rovelli:2014ssa, Finocchiaro:2018hks}, dynamical triangulations~\cite{Ambjorn:2001cv,Gorlich:2013hu,Ambjorn:2013tki,Loll:2019rdj} and simplicial gravity path integrals~\cite{Reisenberger:1997sk,Freidel:1998pt,Baratin:2011hp,Finocchiaro:2018hks}.

The field theoretic nature of GFTs is particularly helpful by allowing a rather straightforward application to quantum gravity problems of all the well-established tools from quantum and statistical field theory. Among them, Landau-Ginzburg mean-field theory and the renormalization group (RG) are important ones for extracting effective macroscopic dynamics from the full quantum many-body description. 
In this context, the hypothesis that we follow is that continuum gravitational physics should be sought for in a condensate phase of a candidate GFT model for $4d$ Lorentzian quantum gravity. This hypothesis, while remaining such, is increasingly supported by growing number of mean-field~\cite{Pithis:2018eaq,Pithis:2019mlv,Marchetti:2021fix,Marchetti:2021xyz} and renormalization group studies~\cite{Benedetti:2015et,BenGeloun:2015ej,Benedetti:2016db,BenGeloun:2016kw,Carrozza:2016tih,Carrozza:2017vkz,BenGeloun:2018ekd,Pithis:2020kio,Pithis:2020sxm,Lahoche:2021nba}, from the more formal perspective. It is also supported by the concrete examples of effective cosmological dynamics obtained in recent years, as we discuss in the following. 

The scenario motivating this hypothesis is that of a transition between a pre-geometric phase, which cannot be interpreted in terms of well-defined spatiotemporal notions, and a geometric one characterized by semi-classical continuum spacetimes~\cite{Oriti:2013jga}, with the latter physics approximately captured in terms of GFT condensates.

Taking seriously this hypothesis, GFT condensate states and their dynamics in relation to cosmology have been intensely studied in recent years~\cite{Gielen:2013kla,Gielen:2013naa,Oriti:2016qtz,Gielen:2016dss,Oriti:2016acw,Pithis:2019tvp}. The effective relational dynamics of the universe volume as a function of scalar field values (used as a clock) is extracted from the mean-field equations of motion of the condensate wavefunction (playing the role of order parameter, since it collectively controls the behavior of microscopic quantum gravity constituents)~\cite{Oriti:2016qtz,Li:2017uao,Gielen:2018fqv}. Based on the GFT formulation~\cite{BenGeloun:2010qkf,Oriti:2016qtz} of the EPRL spin foam model~\cite{Engle:2007wy,Rovelli:2011eq,Perez:2012wv} for Lorentzian quantum gravity in four dimensions, suitably restricted to isotropic configurations\footnote{Many details of the underlying model become irrelevant in this restriction, and the analysis also does not depend on the detailed form of the interaction kernel (where the Lorentzian data are encoded). Thus, the underlying model in these works is referred to as \enquote{EPRL-like}, and the results are understood as having a more general validity.} numerous results have so far been obtained. Most strikingly, the condensate dynamics give rise to quantum corrected Friedmann equations resolving the big bang singularity into a big bounce~\cite{Oriti:2016ueo,Oriti:2016qtz}. Furthermore, studies in~\cite{Gielen:2016uft,Pithis:2016wzf} showed that under quite generic conditions, GFT condensates are dominated by quanta with small triangle areas at late times. Based on these results, it was shown that the quantum geometry described by the condensates isotropizes and classicalizes  dynamically~\cite{Pithis:2016cxg}. Moreover, phenomenological implications of the GFT condensate approach, such as geometric inflation~\cite{deCesare:2016rsf,deCesare:2016axk,Pithis:2016cxg} or emergent dark energy~\cite{Oriti:2021rvm}, have been explored.

The results derived from the EPRL-like GFT model are promising, but the cosmological sector of other viable GFT models has to be studied and results should be compared, to choose between existing models or to identify universal features. 

For instance, next to the EPRL spin foam model, another prominent one is the model by Barrett and Crane (BC)~\cite{Perez:2000ec,Barrett:1999qw}, see also~\cite{DePietri:1999bx,Perez:2000ec,Perez:2000ep,Baratin:2011tx} for its respective GFT formulation. A number of criticisms have been put forward against this model, also in comparison with the EPRL model leading to a decline in attention in recent years. A main point in favor of the EPRL model is that its boundary states are much closer to the ones of canonical LQG, so that it can be considered as its covariant version, since it incorporates the Barbero-Immirzi parameter $\gamma$ of the canonical theory. Second, concerns have been raised about the semi-classical behaviour of the BC model for fixed lattice~\cite{Barrett:2002ur,Kaminski:2013yca,Dittrich:2021kzs}, possibly due to the way geometricity constraints are imposed in the construction (see~\cite{Baratin:2010wi,Baratin:2011tx} for a discussion). Still, on the one hand, none of the criticisms against the BC model is conclusive; on the other hand, the absence of the Barbero-Immirzi parameter is, per se, not a good reason to expect that the model does not capture the correct continuum gravitational physics. In fact, besides the discussion in~\cite{Baratin:2011tx}, recent results in an effective analysis of spin foam amplitudes suggest that both EPRL and BC models could lie in the same continuum universality class and both capture the right semi-classical physics~\cite{Dittrich:2021kzs}. 

This motivates us to look at the effective cosmological dynamics of the BC model. To this aim, we develop (and then use) an extended GFT formulation of the Lorentzian Barrett-Crane model, adapting earlier work in the Riemannian context. In~\cite{Baratin:2011tx} an extension of the Riemannian BC GFT formulation has been introduced, where a dynamical vector, playing the role of normal vector to the tetrahedra corresponding to the GFT quanta, is added to the four $\text{SO}(4)$ group elements representing the quantum geometric data defining the domain of the GFT field. In our Lorentzian context, the additional variable is a timelike normal vector, valued in the upper hyperboloid in Minkowski space. The main advantage is that now simplicity and closure constraints can be imposed in a covariant and commuting fashion, such that the resulting edge spin foam amplitudes are unambiguously defined.  

A brief outline of this article is as follows. 
We set up the extended Lorentzian BC GFT model in Section~\ref{sec:Extended Barrett-Crane Group Field Theory Model} and briefly review how a scalar field is coupled and how operators and the quantum equations of motion are built. The detailed construction of this (extended) Lorentzian BC model is our first result.
In Section~\ref{sec:GFT Condensates}, we introduce the condensate states and specify the domain of the condensate wavefunction for the extended BC GFT model. 
With the diffeomorphism of the domain of the condensate wavefunction with minisuperspace of homogeneous geometries, an important intermediate result is obtained in Section~\ref{subsec:Isomorphism of Domain and Minisuperspace}. 
The effective cosmology arising from the extended BC GFT model is studied and compared with the EPRL-like results in Section~\ref{sec:Evolution of Effective Cosmology}. This is the main focus, and main result of our work.
To conclude, a general discussion of the results, their limitations and possible extensions is given in Section~\ref{sec:Discussion and Conclusion}. 
In Appendices \ref{appendix:Representation Theory of SL2C}, \ref{appendix:Recoupling Theory of SL2C} and \ref{appendix:Spin representation of the extended BC model}, we present the essential notions of $\SL$-representation and -recoupling theory and the novel derivation of the spin representation of a group field in the extended Lorentzian formalism with timelike normal. 

\section{Extended Barrett-Crane group field theory model}\label{sec:Extended Barrett-Crane Group Field Theory Model}

Barrett and Crane uniquely characterized a 4-simplex embedded in Euclidean space $\R^4$~\cite{Barrett:1997gw} and in Minkowski space $\R^{1,3}$~\cite{Barrett:1999qw} in terms bivectors satisfying six conditions\footnote{The so-called non-degeneracy conditions account for non-zero volume of boundary tetrahedra and for the $4$-simplex not to lie in a three-dimensional hyperplane. It is an open issue to implement them in both, spin foam models and GFTs~\cite{Dittrich:2010ey,Baratin:2011tx}, wherefore non-degeneracy conditions are imposed 'by hand' where needed in the remainder.} which form the basis of the Barrett-Crane (BC) spin foam model~\cite{Barrett:1999qw}. In the following, introducing a field-theoretic perspective on these matters, we establish a GFT formulation of the BC spin foam model which goes beyond the original formulation given in~\cite{Perez:2000ec}. Tools from this field-theoretic setting will consequently allow to consider condensates of GFT quanta, leading to GFT condensate cosmology. 

\subsection{Extension of the domain}

In~\cite{Baratin:2011tx}, an extended version of the Euclidean BC GFT model was introduced, resolving the issues of non-covariant and non-commutative imposition of simplicity and closure constraint.\footnote{Ambiguities in older versions of the BC GFT model~\cite{Perez:2000ec,Rovelli:2004tv} led to different edge amplitudes, depending on the order of imposing simplicity and closure. For instance, three different models referred to as BC-A, BC-B and BC-C are discussed in~\cite{Rovelli:2004tv}.} In this article, we work with the extended Lorentzian BC GFT model (which is a generalization of the Euclidean model in~\cite{Baratin:2011tx}), where the domain of the GFT field is extended to include the upper of the two-sheeted 3-hyperboloid $\HH$ defined in Appendix~\ref{appendix:Representation Theory of SL2C}
\begin{equation}
\SL^4\longrightarrow\SL^4\times\HH,
\end{equation}
by introducing $X\in\HH$ as an additional fifth argument, i.e.
\begin{equation}
\varphi(g_v)\longrightarrow\varphi(g_v;X),
\end{equation}
where $g_v\equiv (g_1,g_2,g_3,g_4)$. The timelike vector $X$, interpreted as normal vector to the tetrahedra dual to GFT quanta, enters the action \eqref{eq:divergent BO action} to ensure the correct imposition of geometricity constraints, but drops from the perturbative amplitudes. A more detailed geometric interpretation of the field domain will be given in Sec. \ref{subsec:Isomorphism of Domain and Minisuperspace}.

\begin{figure}[h!]
    \centering
    \includegraphics[scale = 0.2, trim = 60pt 120pt 110pt 30pt, clip]{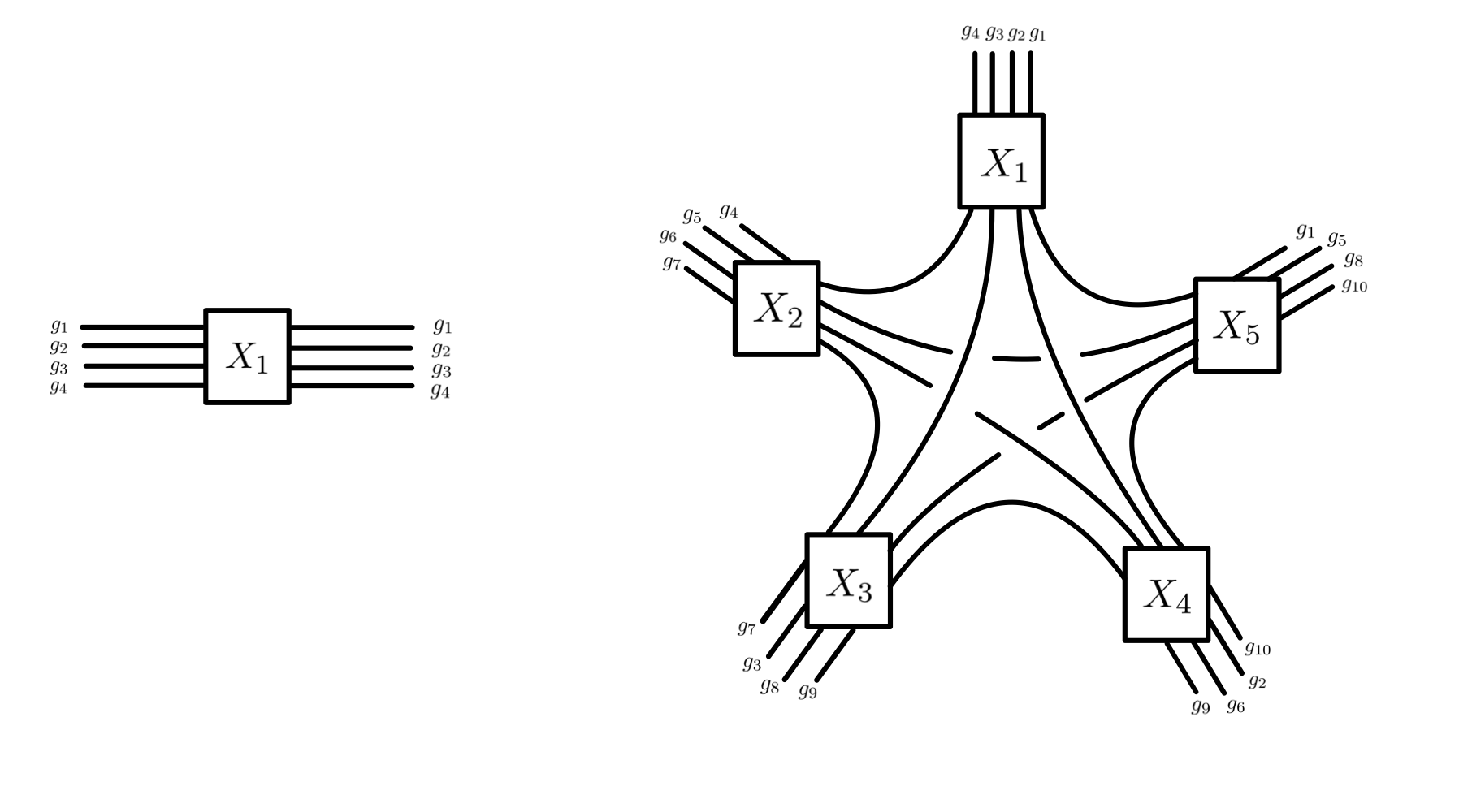}
    \caption{Left panel: The kinetic term of the group field accounts for the gluing of strands and the identification of normal vectors. Right panel: Simplicial interactions lead to a gluing of strands according to a $4$-simplex. The normal vectors in Eq.~\eqref{eq:divergent BO action} are not coupled and therefore integrated over separately, indicated by the box.}
    \label{fig:sketch}
\end{figure}

Analogously to the action of~\cite{Baratin:2011tx} for the Euclidean model, the extended Lorentzian BC GFT model is defined by
\begin{equation}\label{eq:divergent BO action}
\begin{aligned}
& S[\bar{\varphi},\varphi]
=
K + V
=
\int\left[\dd{g}\right]^4\int\dd{X}\bar{\varphi}(g_v;X)\varphi(g_v;X) + \\[7pt]
& +
\frac{\lambda}{5}\int\left[\dd{g}\right]^{10}\int\left[\dd{X}\right]^5 \varphi_{1234}(X_1)\varphi_{4567}(X_2)\varphi_{7389}(X_3)\varphi_{962(10)}(X_4)\varphi_{(10)851}(X_5) +c.c.,
\end{aligned}
\end{equation}
using the notation $\varphi_{1234}(X) \equiv \varphi(g_1, g_2, g_3 ,g_4;X)$, where $\varphi(g_v;X)$ exhibits the following symmetries
\begin{align}
\varphi(g_1,g_2,g_3,g_4;X) &= \varphi(g_1 u_1, g_2 u_2, g_3 u_3, g_4 u_4;X),\quad \forall u_i\in\SU_{X},\label{eq:generalized simplicity}\\[7pt]
\varphi(g_1,g_2,g_3,g_4;X) &= \varphi(g_1 h^{-1}, g_2 h^{-1}, g_3 h^{-1}, g_4 h^{-1}; h\cdot X),\quad \forall h\in\SL,\label{eq:covariance under right}
\end{align}
referred to as simplicity and right-covariance\footnote{The explicit action of $\SL$ on $\HH$ is given in Appendix ~\ref{appendix:Representation Theory of SL2C}.}, the geometric interpretation of which is specified below.\footnote{The interaction term does not incorporate mixed terms of $\varphi$ and $\bar{\varphi}$. As for matrix models~\cite{DiFrancesco:1993cyw}, this allows for the orientability of the simplicial complexes dual to Feynman graphs, see~\cite{Gurau:2011xp} and in particular~\cite{Caravelli:2010nh}.} Crucially, the simplicity constraint turns the model from a GFT description of topological $BF$-theory into one for first-order Palatini gravity. A graphical interpretation of the kinetic and interaction term of the action in Eq.~\eqref{eq:divergent BO action} is given in Fig.~\ref{fig:sketch}.

Due to the projector nature of the operators acting on the GFT field in the action, and encoding the imposition of the constraints, there are no ambiguities in the definition of the action and in the resulting Feynman (spin foam) amplitudes, since any other choice of imposition (e.g. ordering), would be equivalent to the given one. See also~\cite{Baratin:2011tx}. Notice, that the timelike normals are simply identified in the kinetic kernel and enter the vertex term non-dynamically, i.e. without any coupling across fields. As a result, they do not enter the resulting Feynman amplitudes, and only serve as auxiliary variables that contain information about the embedding of tetrahedra. Clearly, action \eqref{eq:divergent BO action} is divergent as written, due to the non-compactness of the Lorentz group. A proper regularization will be given shortly.

Different representations of the GFT field are possible, one of which is the Lie algebra representation (also called flux or metric representation), obtained by the non-commutative Fourier transform~\cite{Baratin:2010wi,Guedes:2013vi,Oriti:2018bwr}
\begin{equation}
\tilde{\varphi}_X(B_1, B_2, B_3, B_4)
=
\int\left[\dd{g}\right]^4 e_{g_1}(B_1)e_{g_2}(B_2)e_{g_3}(B_3)e_{g_4}(B_4)\varphi(g_1,g_2,g_3,g_4;X),
\end{equation}
where $e_{g_i}$ are non-commutative plane waves and $B_i\in\spl$ are interpreted as bivectors of triangles, the norm of which corresponds to the triangle area.\footnote{This interpretation is justified by the isomorphism $\spl\cong\bigwedge^2\R^{1,3}$~\cite{Pereira:2010wzm}, where the electric and magnetic components of bivectors $K^a\defeq B^{a0}$, $L^a\defeq \frac{1}{2}\tensor{\varepsilon}{^a_{bc}}B^{bc}$ are given by the generators of boosts and rotations as defined in Appendix \ref{appendix:Representation Theory of SL2C}. Since the Minkowski metric on $\R^{1,3}$ canonically extends to $\bigwedge^2\R^{1,3}$, we can determine bivectors to be timelike, spacelike or null. Moreover, the Hodge star operator $*$ is introduced, which interchanges spacelike and timelike bivectors~\cite{Barrett:1999qw} or equivalently, interchanges electric and magnetic components.} In this representation, simplicity \eqref{eq:generalized simplicity} and right-covariance \eqref{eq:covariance under right} acquire a clear geometric interpretation. Bivectors, associated to triangles of tetrahedra $(t\in\tau)$, are forced to be simple with respect to the timelike normal $X$, i.e.
\begin{equation}\label{eq:simplicity on bivectors}
X_A (*B)^{AB} =  0\quad\Rightarrow\quad 
\exists\; v_t,w_t\in\R^{1,3}:B_t=v_t\wedge w_t,
\end{equation}
where $A,B$ are Lorentz indices. Furthermore, the bivectors close upon integration of the timelike normal 
\begin{equation}
\sum_{t\in\tau}B_t = 0.
\end{equation}
In Section \ref{subsec:Isomorphism of Domain and Minisuperspace} we are going to show in detail how bivectors determine the local $3$-geometry, for which an explicit mapping is given in Eq.~\eqref{eq:map to metric coefficients}. 
In addition to a clear geometric interpretation of the field configurations, the flux representation makes the connection of GFTs and simplicial path integrals explicit~\cite{Oriti:2011jm, Finocchiaro:2018hks}. 

In the following section, we introduce the spin representation of the GFT field, re-write the action \eqref{eq:divergent BO action} in terms of representation labels of $\SL$ and provide a regularization.

\subsection{Spin Representation and Regularization }\label{subsec:Spin Representation and Regularization of Group Field}

We are interested in the expansion of functions on the extended domain $\SL^4\times\HH$. 
For Euclidean signature,~\cite{Baratin:2011tx} provides a detailed description of the spin representation of the group field, relying on the structure of the group $\text{SO}(4)$. 
The expansion in terms of $\SL$-representation labels $(\rho,\nu)\in\R\times\mathbb{Z}/2$ of the Lorentzian extended BC model is derived in Appendix~\ref{appendix:Spin representation of the extended BC model}, which is important for the ensuing condensate cosmology analysis.

Following the derivation of Appendix~\ref{appendix:Spin representation of the extended BC model}, we impose simplicity on a right-covariant function with timelike normal $X\equiv[a]$ defined by Eq.~\eqref{eq:group field with right covariance}, yielding
\begin{equation}\label{eq:group field with right covariance, simplicity}
\varphi(g_v;[a])
=
\left[\prod_i \int\dd{\rho_i}4\rho_i^2\sum_{j_i m_i}\right]\varphi^{\rho_i0,00}_{j_i m_i}\prod_{i} D^{(\rho_i,0)}_{j_i m_i 00}(g_ia),
\end{equation}
where we henceforth set $\varphi^{\rho_i 0,00}_{j_i m_i} \equiv \varphi^{\rho_i}_{j_i m_i}$ for notational ease. The $j_i\in\{\abs{\nu_i},\abs{\nu_i}+1,...\}$ and $m_i\in\{-j_i,...,+j_i\}$ are labels of basis vectors in $\SL$-representation spaces. In terms of representation labels, the simplicity condition \eqref{eq:simplicity on bivectors} implies that the second Casimir $\cas_2 = B_t\cdot *B_t$ defined in Eq.~\eqref{eq:definition of cas2} with eigenvalues $\rho_t\cdot \nu_t$ vanishes. The first solution $\nu_i = 0$, which is also the one we restrict ourselves to for the remainder of this work, is obtained by modding out the rotational subgroup $\SU$ yielding the homogeneous space $\SL/\SU$, the bivectors of which are spacelike.\footnote{As our aim is to describe homogeneous cosmologies by condensates of tetrahedra forming spacelike hypersurfaces, we do not consider the case of timelike tetrahedra here. However, note that the other solution of the equation, namely $\rho_i = 0$, is obtained by modding out the subgroup SU$(1,1)$, i.e. by considering the homogeneous space $\SL/$SU$(1,1)$, the normal of which is spacelike and the bivectors are timelike or spacelike, see~\cite{Perez:2000ep} for a detailed description.}

In the interaction term of Eq.~\eqref{eq:divergent BO action}, the GFT fields enter with their timelike normal integrated over separately, giving rise to the Barrett-Crane intertwiners~\cite{Barrett:1999qw,Oriti:2003wf}
\begin{equation}\label{eq:definition of BC intertwiner}
B^{\rho_i}_{j_i m_i} 
\equiv
B^{\rho_1 \rho_2 \rho_3 \rho_4}_{j_1 m_1 j_2 m_2 j_3 m_3 j_4 m_4}
\defeq
\int\limits_{\HH}\dd{X}\prod_{i=1}^4 D^{(\rho_i,0)}_{j_i m_i 00}(X).
\end{equation}
As a consequence, the GFT field assumes the form
\begin{equation}\label{eq:group field with all right covariance, simplicity and integration over normal}
\int\limits_{\HH}\dd{X}\varphi(g_v;X)
=
\left[\prod_i \int\dd{\rho_i}4\rho_i^2\sum_{j_i m_i l_i n_i}\right]\varphi^{\rho_i}_{j_i m_i}B^{\rho_i}_{l_i n_i}\prod_{i}D^{(\rho_i,0)}_{j_i m_i l_i n_i}(g_i).
\end{equation}

Altogether, the GFT action in terms of representation labels is computed by applying Eq.~\eqref{eq:group field with right covariance, simplicity} and Eq.~\eqref{eq:group field with all right covariance, simplicity and integration over normal} 
\begin{equation}\label{eq:BO action in spin representation}
\begin{aligned}
S 
& = 
\left[\prod_i \int\dd{\rho_i}4\rho_i^2\sum_{j_i m_i}\right]\bar{\varphi}^{\rho_i}_{j_i m_i}\varphi^{\rho_i}_{j_i m_i}\int\limits_{\HH}\dd{X}+\frac{\lambda}{5}\left[\prod_{a=1}^{10}\int\dd{\rho_a}4\rho_a^2\sum_{j_a m_a}\right]\times\\[7pt]
& \times
\left(\prod_{a=1}^{10}(-1)^{-j_a-m_a}\right)\{10\rho\}_{\text{BC}}\varphi^{\rho_1 \rho_2 \rho_3 \rho_4}_{j_1 m_1 j_2 m_2 j_3 m_3 j_4 m_4}\varphi^{\rho_4 \rho_5 \rho_6 \rho_7}_{j_4 -m_4 j_5 m_5 j_6 m_6 j_7 m_7}\times\\[7pt]
& \times
\varphi^{\rho_7 \rho_3 \rho_8 \rho_9}_{j_7 -m_7 j_3 -m_3 j_8 m_8 j_9 m_9}\varphi^{\rho_9 \rho_6 \rho_2 \rho_{10}}_{j_9 -m_9 j_6 -m_6 j_2 -m_2 j_{10} m_{10}}\varphi^{\rho_{10} \rho_8 \rho_5 \rho_1}_{j_{10} -m_{10} j_8 -m_8 j_5 -m_5 j_1 -m_1}+c.c,
\end{aligned}
\end{equation}
where we observe the BC $\{10\rho\}$-symbol defined in Eq.~\eqref{eq:definition of BC 10p symbol} and a redundant integration over the timelike normal. As presented in~\cite{Perez:2000ec}, the BC $\{10\rho\}$-symbol can be brought to the form given in Eq.~\eqref{eq:integral form of 10p}. After a successive re-definition of variables $h_i$ in Eq.~\eqref{eq:integral form of 10p}, one integration over $\SL$ is redundant leading to a divergent factor, formally denoted by vol$(\SL)$. We then define the regulated BC $\{10\rho\}$-symbol
\begin{equation}\label{eq:definition of regulated BC 10p}
\{10\rho\}_{\text{BC}}
\eqdef
\{10\rho\}_{\text{BC}}^{\text{reg}}\cdot \text{vol}(\SL).
\end{equation}
On the other hand, using the Cartan decomposition of the Haar measure on $\SL$, given in Eq.~\eqref{eq:Cartan decomposition of Haar measure}, it is clear that only the boost part of a redundant $\SL$-integration, i.e. the hyperbolic part, is divergent while the rotation part yields a factor of one. Therefore, redundant integrations over $\HH$ and $\SL$ diverge equally and kinetic and vertex terms contain the same degree of divergence, allowing for a factorization of vol$(\SL)$ in front of the regulated action. Henceforth, we work with the regulated action from now on and drop any indication of regularization for brevity.

Equation~\eqref{eq:BO action in spin representation} defines a variant of the Barrett-Crane spin foam model~\cite{Perez:2000ec,Oriti:2003wf,Barrett:1999qw}, which serves as a self-consistency check for the spin representation we have constructed. In particular, the vertex amplitude obtained from the interaction kernel agrees with the one obtained in~\cite{Perez:2000ec}.

Corresponding to the (regularized version of the) action~\eqref{eq:BO action in spin representation}, the equations of motion for the GFT field $\varphi^{\rho_i}_{j_i m_i}$ in spin representation are obtained via the variation
\begin{equation}
\fdv{S[\varphi,\bar{\varphi}]}{\bar{\varphi}^{\rho_1 \rho_2\rho_3\rho_4}_{j_1 m_1 j_2 m_2 j_3 m_3 j_4 m_4}} = 0,
\end{equation}
leading to
\begin{equation}
\begin{aligned}
0 & = \left(\prod_{i=1}^44\rho_i^2\right)\Bigg{\{}\varphi^{\rho_i}_{j_i m_i}+\lambda\left[\prod_{a=5}^{10}\int\dd{\rho_a}4\rho_a^2\sum_{j_a m_a}\right]\{10\rho\}_{\text{BC}}\bar{\varphi}^{\rho_4 \rho_5 \rho_6 \rho_7}_{j_4 -m_4 j_5 m_5 j_6 m_6 j_7 m_7}\times\\[7pt]
& \times
\bar{\varphi}^{\rho_7 \rho_3 \rho_8 \rho_9}_{j_7 -m_7 j_3 -m_3 j_8 m_8 j_9 m_9}\bar{\varphi}^{\rho_9 \rho_6 \rho_2 \rho_{10}}_{j_9 -m_9 j_6 -m_6 j_2 -m_2 j_{10} m_{10}}\bar{\varphi}^{\rho_{10} \rho_8 \rho_5 \rho_1}_{j_{10} -m_{10} j_8 -m_8 j_5 -m_5 j_1 -m_1}\Bigg{\}}.
\end{aligned}
\end{equation}

Instead of averaging over the timelike normal, a different route can be taken by fixing the timelike normal to some value $X_0\in\HH$, e.g. to time-gauge $X_0 = [e]$, using the right covariance condition. As a consequence, the right $\SL$ action is reduced to an $\SU_{X_0}$ action and the right-covariance condition trivializes entirely (the leftover diagonal $SU(2)_{X_0}$ rotation acts trivially). For a well-defined geometric interpretation, an additional closure condition is required, yielding again a Barrett-Crane intertwiner. Although a comparison to models based on the $3+1$ splitting of spacetime, such as the canonical formulation of Palatini~\cite{Ashtekar1991} or ADM gravity~\cite{Thiemann:2007pyv}, is more illustrative in this formulation, the required closure condition is ad hoc and needs to be imposed artificially. In contrast, the extended Lorentzian BC GFT model represents a covariant treatment in the sense that there is no preferred foliation. Furthermore, the closure arises naturally upon timelike normal integration and does not need to be introduced 'by hand'.

In the following section we briefly specify the quantization of the GFT field. Thereafter, we couple additional degrees of freedom corresponding, at the level of the discrete path integral dual to the Feynman amplitudes of the model, to a discretized real free massless scalar field ~\cite{Li:2017uao}. These additional degrees of freedom will later be used to define relational observables and a proper notion of cosmological evolution.

\subsection{Promoting GFT fields to operators}

Following~\cite{Oriti:2013aqa} and adapting it to the case of the BC model, we quantize the GFT fields by promoting them to operators acting on the Fock space
\begin{equation}\label{eq:definition of Fock space}
\mathcal{F} 
\defeq
\bigoplus_{V = 0}^\infty \text{sym}\left(\mathcal{H}_v^{(1)}\otimes...\otimes\mathcal{H}_v^{(V)}\right),
\end{equation}
and specifying their commutation relations, assuming bosonic statistics\footnote{For explorations on other statistics, see~\cite{Gurau:2011xp} and~\cite{Girelli:2010ct,Baratin:2011tg}.}
\begin{equation}\label{eq:commutation relations of group field}
\begin{aligned}
\comm{\hat{\varphi}(g_v;X)}{\hat{\varphi}^{\dagger}(g_w;X')} & = \one(g_v;X,g_w;X'),\\[7pt]
\comm{\hat{\varphi}(g_v;X)}{\hat{\varphi}(g_w;X)} &= \comm{\hat{\varphi}^{\dagger}(g_v;X)}{\hat{\varphi}^{\dagger}(g_w;X)} = 0,
\end{aligned}
\end{equation}
where $\one$ is the identity on the space $L^2\left(\SL^4\times\HH\right)$ respecting simplicity and right-covariance. As in standard quantum field theory, the definition of a vacuum state $\ket{\emptyset}$ follows immediately
\begin{equation}
\hat{\varphi}(g_v;X)\ket{\emptyset} = 0,\quad \forall (g_v;X)\in \SL^4\times\HH,
\end{equation}
interpreted as the state of no space.

The single-particle Hilbert space, or equivalently the Hilbert space of an atom of space, $\mathcal{H}_v$ is defined as an $L^2$-space on the quotient defined by the two conditions of Eq.~\eqref{eq:generalized simplicity} and Eq.~\eqref{eq:covariance under right}. Its interpretation and use in the cosmological application, will be presented in Secs.~\ref{subsec:Domain of Condensate Wavefunction} and~\ref{subsec:Isomorphism of Domain and Minisuperspace}. 

\subsection{Coupling to a scalar field}\label{subsec:coupling to a scalar field}

As it is common practice in the quantum gravity literature~\cite{Oriti:2016qtz,Pithis:2019tvp,Marchetti:2020umh,Giesel:2012rb,Ashtekar:2011ni}, we introduce a real massless free scalar field $\phi$ serving as a relational clock to describe cosmological evolution with respect to an internal degree of freedom.

Following~\cite{Oriti:2016qtz,Li:2017uao}, the clock field $\phi$ is coupled to the GFT model by further extending the domain of the GFT field, leading to
\begin{equation}
\varphi(g_v;X) \longrightarrow \varphi(g_v;X;\phi),
\end{equation}
and extending the kernels $\mathcal{K}(g_v,w_w)$ and $\mathcal{V}(g_{v_1},...,g_{v_5})$ in the action \eqref{eq:divergent BO action}, to
\begin{align}
\mathcal{K}(g_v,g_w) &\longrightarrow \mathcal{K}\left(g_v,g_w;(\phi_v-\phi_w)^2\right),\label{eq:extension of kinetic kernel}\\[7pt]
\mathcal{V}(g_{v_a}) &\longrightarrow\mathcal{V}(g_{v_a}).
\end{align}
We assume that there exists a series expansion of the kinetic kernel $\mathcal{K}\left(g_v,g_w;(\phi_v-\phi_w)^2\right)$
\begin{equation}\label{eq:expansion of kinetic kernel}
\mathcal{K}\left(g_v,g_w;(\phi-\phi_0)^2\right)
=
\sum_{n=0}^{\infty}\frac{\mathcal{K}_{(2n)}(g_v,g_w)}{(2n)!}(\phi-\phi_0)^{2n} .
\end{equation}
This expansion will be used for the derivation of an effective action in Section~\ref{subsec:Spin Representation of the Action and Regularization}.

\subsection{2nd quantized operators}\label{subsec:2nd Quantized Operators}

Since the GFT field operator $\hat{\varphi}$ satisfies the commutation relations \eqref{eq:commutation relations of group field}, the construction of many-body operators lies at hand. For the application to homogeneous cosmologies in Secion~\ref{subsec:Coherent Peaked Condensate States}, one-body operators suffice for a first analysis.\footnote{Similar to classical homogeneous and isotropic cosmology, where the scale factor (or equivalently the volume) encode the whole dynamical information, the dynamics of coherent condensate cosmology are determined by the one-body volume operator \eqref{eq:definition of volume operator}.}  These have the general form of suitable kernels convoluted with the (number) operator $\hat{\varphi}^{\dagger}(g_v;X;\phi)\hat{\varphi}(g_v;X;\phi)$. The three most relevant examples are the number operator $\hat{N}$ itself, the volume operator $\hat{V}$ and the scalar field momentum operator $\hat{\pi}_{\phi}$.

Clearly, the number operator is defined as
\begin{equation}\label{eq:definition of relational number operator}
\hat{N}
=
\int\dd{\phi}\int\left[\dd{g}\right]^3\int\dd{X}\hat{\varphi}^{\dagger}(g_v;X;\phi)\hat{\varphi}(g_v;X;\phi).
\end{equation}

Following~\cite{Barrett:1997gw}, the volume of a tetrahedron is characterized in terms of bivectors via 
\begin{equation}
V(B_1,B_2,B_3)\propto\sqrt{\abs{\varepsilon^{ijk}\tr\left(B_i B_j B_k\right)}} = \sqrt{\abs{\varepsilon^{ijk}\delta_{BC}\delta_{DE}\delta_{FA}B_i^{AB}B_j^{CD}B_k^{EF}}},
\end{equation}
yielding the volume operator in bivector (Lie algebra) representation
\begin{equation}\label{eq:definition of volume operator}
\hat{V}=\int\dd{\phi}\int\left[\dd{B}\right]^3\int\dd{X}\hat{\tilde{\varphi}}^{\dagger}_X(B_v;\phi)\star V(B_1,B_2,B_3)\hat{\tilde{\varphi}}_X(B_v;\phi).
\end{equation}
In fact, the only property which is relevant for the derivation of the effective cosmological evolution in Sec.~\ref{sec:Evolution of Effective Cosmology}, is that the kernel $V(B_1,B_2,B_3)$ scales in the monochromatic case, i.e. $B_1 = B_2 = B_3 \equiv B$, as
\begin{equation}
V(\{B\})\sim\rho^{3/2},
\end{equation}
where $\rho$ is the continuous representation label of $\SL$.\footnote{Given the spectrum of the area of the individual faces of a BC tetrahedron, which is discussed below in Eq.~\eqref{eq:proportionality BC area}, it is reasonable to assume that for dimensional reasons this is the only consistent scaling in the monochromatic case. A detailed discussion of the diagonalization of this operator in the general case is left to future investigations and can most likely be accomplished in close analogy with results of LQG~\cite{Rovelli:1994ge,Barbieri:1997ks,Brunnemann:2004xi,Ding:2009jq}.}

The scalar field momentum operator, entering the definition of the energy density in Section~\ref{subsec:Emergent Quantum Bounce}, is defined by
\begin{equation}\label{eq:clock conjugate momentum operator}
\hat{\pi}_{\phi}
=
\frac{\hbar}{2i}\int\dd{\phi}\int\left[\dd{g}\right]^3\int\dd{X}\left[\hat{\varphi}^{\dagger}(g_v;X;\phi)\pdv{}{\phi}\hat{\varphi}(g_v;X;\phi) - \left(\pdv{}{\phi}\hat{\varphi}^{\dagger}(g_v;X;\phi)\right)\hat{\varphi}(g_v;X;\phi)\right].
\end{equation}

Similar to the regularization scheme above we drop one $\SL$-integration in Eq.~\eqref{eq:definition of relational number operator} and Eq.~\eqref{eq:clock conjugate momentum operator} and one $\spl$-integration in  Eq.~\eqref{eq:definition of volume operator}.

We refrain from using 'relational operators' introduced in~\cite{Oriti:2016qtz} that are infinitely peaked on single scalar field values $\phi$, defined as the integrands of the full convolution. Operators constructed this way are mathematically ill-defined and physically relevant quantities such as the scalar field momentum operator are not self-adjoint and subject to arbitrarily large quantum fluctuations~\cite{Marchetti:2020umh,Marchetti:2020qsq}. A way out of this difficulty could be a suitable smearing procedure. Instead, relational evolution will be captured in an effective sense only, in terms of expectation values of the above fundamental operators in coherent condensate states peaked on specific values of the scalar field data ~\cite{Marchetti:2020umh}, that we are going to introduce in Section~\ref{subsec:Coherent Peaked Condensate States}.

\subsection{Quantum equations of motion}\label{subsec:Quantum Equations of Motion}

The full quantum dynamics can be encoded in the non-perturbative quantum equations of motion of the GFT field operator
\begin{equation}\label{eq:operator quantum equations of motion}
\widehat{\frac{\delta S[\varphi,\bar{\varphi}]}{\delta\bar{\varphi}(g_v;X;\phi)}}\ket{\Psi} = 0 , 
\end{equation} 
or equivalently, determined by the infinite tower of Schwinger-Dyson equations~\cite{Gielen:2013naa}
\begin{equation}\label{eq:Schwinger-Dyson equations}
0
=
\expval{\fdv{\mathcal{O}[\varphi,\bar{\varphi}]}{\bar{\varphi}(g_v;\phi)}-\mathcal{O}[\varphi,\bar{\varphi}]\fdv{S[\varphi,\bar{\varphi}]}{\bar{\varphi}(g_v,\phi)}},
\end{equation}
where $\mathcal{O}[\varphi,\bar{\varphi}]$ is any polynomially bounded functional. For $\mathcal{O}[\varphi,\bar{\varphi}] = 1$, one obtains the simplest equation of \eqref{eq:Schwinger-Dyson equations}. We will only consider the approximate dynamics captured by coherent condensate states solutions of such simplest equation.

\section{GFT coherent condensate states}\label{sec:GFT Condensates}

GFT coherent condensate states have proven to be sufficient to capture relevant cosmological phenomena despite their simple form, see~\cite{Pithis:2019tvp} for a review. They assign the same condensate wavefunction to every GFT quantum, and this can be seen as representing a microscopic quantum analogue of the notion of homogeneity. Most importantly, the provide the simplest approximation to the full continuum quantum dynamics, in the form of a mean field hydrodynamics, with the mean field corresponding to the condensate wavefunction. This corresponds to the Gross-Pitaevskii hydrodynamics in the theory of superfluids~\cite{pitaevskii2016bose}. To obtain from such hydrodynamic equations a geometrically more transparent cosmological evolution in terms of relational observables, we will specialize to coherent peaked states (CPS) introduced in~\cite{Marchetti:2020umh} for the EPRL-like model and adapt them to the case of the BC model in Sections~\ref{subsec:Coherent Peaked Condensate States} and~\ref{subsec:Spin Representation of the Action and Regularization} in the following. Representing novel results, we clarify the relation between the domain of the condensate wavefunction and minisuperspace and derive the spin representations thereof in Sections~\ref{subsec:Domain of Condensate Wavefunction},~\ref{subsec:Isomorphism of Domain and Minisuperspace} and~\ref{subsec:Spin Representation of Condensate Wavefunction}, respectively.

\subsection{Coherent peaked condensate states}\label{subsec:Coherent Peaked Condensate States} 

According to~\cite{Marchetti:2020umh}, we introduce CPS in order to solve some difficulties with the relational observables used in~\cite{Oriti:2016qtz}. In such states, the GFT quanta are sharply, but not infinitely, peaked around a fixed scalar field value $\phi_0$.

Following~\cite{Marchetti:2020umh}, the coherent peaked state $\ket{\sigma_{\epsilon};\phi_0;\pi_0}$ is defined as
\begin{equation}\label{eq:definition of CPS}
\ket{\sigma_{\epsilon};\phi_0,\pi_0}
\defeq
\e^{-\frac{\norm{\sigma_{\epsilon}}^2}{2}}\exp\left(\int\dd{\phi}\int\left[\dd{g}\right]^2\int\dd{X}\sigma_{\epsilon}(g_v;X;\phi\vert\phi_0,\pi_0)\hat{\varphi}^{\dagger}(g_v;X;\phi)\right)\ket{\emptyset},
\end{equation}
where
\begin{equation}\label{eq:norm of condensate wavefunction}
\norm{\sigma_{\epsilon}}^2 
=
\int\dd{\phi}\int\left[\dd{g}\right]^2\int\dd{X} \bar{\sigma}_{\epsilon}(g_v;X;\phi\vert \phi_0,\pi_0)\sigma(g_v;X;\phi\vert \phi_0,\pi_0),
\end{equation}
and the exponential factor $\mathrm{exp}(-\norm{\sigma_{\epsilon}}^2/2)$ accounts for normalization, i.e. $\braket{\sigma_{\epsilon}} = 1$. The condensate wavefunction $\sigma_{\epsilon}(g_v;X;\phi\vert\phi_0,\pi_0)$ encodes a finite peaking on scalar field values through the decomposition into a peaking function $\eta_{\epsilon}$ and a reduced condensate wavefunction $\sigma$
\begin{equation}
\sigma_{\epsilon}(g_v;X;\phi\vert\phi_0,\pi_0)
=
\eta_{\epsilon}(\phi-\phi_0,\pi_0)\sigma(g_v;X;\phi),
\end{equation}
where we henceforth choose the peaking function to be Gaussian
\begin{equation}
\eta_{\epsilon}(\phi-\phi_0,\pi_0)
=
\mathcal{N}_{\epsilon}\exp(-\frac{(\phi-\phi_0)^2}{2\epsilon})\exp(i\pi_0(\phi-\phi_0)).
\end{equation}
While $\epsilon$ is clearly interpreted as peaking parameter, $\pi_0$ is an additional parameter that needs to be introduced to ensure finite quantum fluctuations of the scalar field momentum~\cite{Marchetti:2020umh}. The peaking property requires $\epsilon\ll 1$. Second, $\epsilon\pi_0^2\gg 1$ is required for the momentum of the scalar field to have small quantum fluctuations, at least in some phase of the effective dynamics. A clock satisfying these two condition is referred to as a \textit{good} clock~\cite{Marchetti:2020umh}. 

Note that, in the (singular) limit $\epsilon\rightarrow 0$, $\sigma_{\epsilon}$ one recovers the  results of~\cite{Oriti:2016qtz}, with the scalar field momentum is however subject to arbitrarily large quantum fluctuations.

Assuming invariance of the state under the symmetries of $\varphi$ implies that the reduced condensate wavefunction $\sigma(g_v;X;\phi)$ carries the same symmetries as the GFT field, given by Eq.~\eqref{eq:generalized simplicity} and Eq.~\eqref{eq:covariance under right}. Notice that in both, Eq.~\eqref{eq:definition of CPS} and Eq.~\eqref{eq:norm of condensate wavefunction}, we dropped two $\SL$-integrations for regularization, anticipating the invariances of $\sigma$ which are specified in Section~\ref{subsec:Domain of Condensate Wavefunction}.

Following the commutation relations \eqref{eq:commutation relations of group field}, the coherent peaked state is an eigenstate of the annihilation operator
\begin{equation}
\hat{\varphi}(g_v;X;\phi)\ket{\sigma_{\epsilon};\phi_0,\pi_0}
=
\sigma_{\epsilon}(g_v;X;\phi\vert\phi_0,\pi_0)\ket{\sigma_{\epsilon};\phi_0,\pi_0}.
\end{equation}
Therefore, the expectation value of the group field with respect to the CPS is
\begin{equation}
\expval{\hat{\varphi}(g_v;X;\phi)}{\sigma_{\epsilon};\phi_0,\pi_0}
=
\sigma_{\epsilon}(g_v;X;\phi\vert\phi_0,\pi_0).
\end{equation}
For this reason, the condensate wavefunction is also referred to as mean-field in analogy to statistical physics, where in the mean-field approximation, quantum operators are replaced by their expectation values.

First, the expectation value of the particle number $N$ is controlled by the reduced condensate wavefunction
\begin{equation}\label{eq:expectation value of number operator}
\expval{\hat{N}}{\sigma_{\epsilon};\phi_0;\pi_0}
=
\norm{\sigma_{\epsilon}}^2
\approx
\int\left[\dd{g}\right]^2\int\dd{X}\abs{\sigma(g_v;X;\phi_0)}^2
\equiv N(\phi_0),
\end{equation}
where we applied a lowest-order saddlepoint approximation~\cite{Marchetti:2020umh}.
For a reasonable interpretation, we require that the reduced condensate wavefunction lives in an $L^2$-space. Therefore, although the coherent state contains terms of infinite quanta, the expectation value of the number operator remains finite. In other words, we do not leave the Fock space structure in a thermodynamic limit but work with Fock coherent states, as opposed to the non-Fock coherent states used in quantum optics~\cite{Honegger1990}.

Moreover, $\ket{\sigma_{\epsilon}}$ is not an exact solution of the quantum equations of motion \eqref{eq:operator quantum equations of motion}, where the error made by the approximation is controlled by the particle number~\cite{Oriti:2016qtz}.

Additionally, the CPS approximation is only compatible with small GFT interactions. From a conceptual point of view, this is in agreement with the product nature of condensate states which therefore do not contain any connectivity information.

Similar to condensed matter physics, where the approximation by Gross-Pitaevskii states already captures enough information to describe the fluid-dynamic behavior of Bose-Einstein condensates~\cite{pitaevskii2016bose}, we expect that $\sigma_{\epsilon}(g_v;X;\phi\vert\phi_0,\pi_0)$ already approximates quantum cosmological evolution well under the above conditions. 

\subsection{Domain of Reduced Condensate Wavefunction}\label{subsec:Domain of Condensate Wavefunction}

The coherent condensate wavefunction, as well as a generic mean field of our GFT model, can be graphically represented just like the individual quanta of the same GFT, via 4-valent open Lorentz spin networks consisting of one vertex and four edges (dual to a tetrahedron and its triangles). On the side of the edges, labelled with representations $(\rho_i,0)$, that meet at the vertex dual to the tetrahedron, we already imposed simplicity and right-covariance, yielding the BC intertwiner. Outward pointing ends of the edges carry non-gauge invariant indices $(j_i,m_i)$ from the canonical basis in the $\rho_i$ Lorentz representation. Capturing the geometric data only (of the dual tetrahedron as well as, in the cosmological interpretation, of a homogeneous universe) requires invariance under the action of the Lorentz group on these open ends. In the GFT condensate cosmology based on EPRL-like models~\cite{Gielen:2014ila,Oriti:2016qtz,deCesare:2017ynn}, this is accomplished by an additional adjoint $\SU$ invariance of the condensate wavefunction, which in Lie algebra variables, is given by the invariance under simultaneous rotations of bivectors. Imposition via group averaging corresponds to averaging over all possible embeddings of tetrahedra, or equivalently, to imposing invariance with respect to local frame rotations. 

Following~\cite{Gielen:2013naa}, Lorentz transformations act on bivectors via the adjoint action of $\SL$. Thus, our open 4-valent Lorentz spin network vertices are effectively closed in a covariant fashion by imposing adjoint-covariance
\begin{equation}\label{eq:covariance under adjoint}
\sigma(g_1, g_2, g_3, g_4;X;\phi)
=
\sigma(hg_1 h^{-1}, hg_2 h^{-1}, hg_3 h^{-1}, hg_4 h^{-1};h\cdot X;\phi),
\end{equation}
similar to the right-covariance \eqref{eq:covariance under right}. Applying the non-commutative Fourier transform on $\SL$ introduced in~\cite{Oriti:2018bwr} and using the transformation property of the plane waves
\begin{equation}\label{eq:adjoint on group is adoint on bivectors}
e_{hg_ih^{-1}}(B_i)
=
e_{g_i}(h^{-1} B_i h),\quad\forall h\in\SL,
\end{equation}
adjoint-covariance can in fact be easily seen to translate to covariance under frame rotations of the bivectors
\begin{equation}
\tilde{\sigma}_X(B_1, B_2, B_3, B_4;\phi)
=
\tilde{\sigma}_{h\cdot X}(h^{-1}B_1 h, h^{-1}B_2 h, h^{-1}B_3 h, h^{-1}B_4 h;\phi).
\end{equation}
Imposing this by group averaging is interpreted as averaging over every configuration involving a preferred hypersurface normal. In this sense, any notion of preferred spacetime foliation (and embedding of tetrahedra in spatial hypersurfaces, in a discrete gravity perspective) is integrated out, resulting in dynamical states that only carry proper geometric information. Consequently, the domain of the reduced condensate wavefunction on the level of bivectors after integrating out the timelike normal can be expected to match the configuration space of classical spatially homogeneous cosmologies. We prove this in the next subsection.

\subsection{Diffeomorphism of condensate domain and minisuperspace}\label{subsec:Isomorphism of Domain and Minisuperspace}

In classical GR, the configuration space of homogeneous spatial hypersurfaces is called minisuperspace. Simply speaking, every point in a homogeneous space looks the same and therefore, the spatial metric field is fully characterized by determining it at a single point. Consequently, the generically infinite-dimensional spatial configuration space of GR reduces to a finite-dimensional one. Since the spatial metric at a point carries nine degrees of freedom but exhibits a rotational symmetry, minisuperspace is clearly given by $\textrm{GL}(3)/\textrm{O}(3)$~\cite{Giulini:2009np}.

We will now show that the domain $\mathfrak{D}$ of the reduced condensate wavefunction in bivector variables\footnote{It is important to stress that, in our analysis, we will consider the space of bivectors only as a vector space, neglecting its non-commutative aspects as a Lie algebra and entering the non-commutative Fourier transform relating them to group variables.} after integrating over the timelike normal is diffeomorphic to minisuperspace. Combined with integration over the normal, adjoint-covariance translates to adjoint-invariance, which is clear from the following consideration
\begin{equation}\label{eq:adjoint covariance becomes adjoint invariance}
\begin{aligned}
& \int\limits_{\HH}\dd{X}\sigma(g_v;X;\phi)
=
\int\dd{X}\sigma(hg_v h^{-1};h\cdot X;\phi)
\overset{X' = h\cdot X}{=}
\int\dd{\left(h^{-1}\cdot X'\right)}\sigma(hg_v h^{-1};X';\phi)\\[7pt]
=
& \int\dd{X'}\sigma(hg_v h^{-1};X';\phi),
\end{aligned}
\end{equation}
where we have used the invariance of the Haar measure on the quotient space $\dd{\left(h\cdot X\right)} = \dd{X}$ for all $h\in\SL$. Similar to Eq.~\eqref{eq:adjoint covariance becomes adjoint invariance}, one can show that right-covariance, as defined in Eq.~\eqref{eq:covariance under right}, reduces to right-invariance under integration over the normal.

As a consequence, the unconstrained domain $\SL^4$ under integration over the timelike normal reduces to $\SL^3$ after imposing closure. Imposing simplicity, yields three factors of the hyperbolic space $\HH$ or equivalently three copies of the Lie group $\SL/\SU \cong \Hom$.\footnote{$\Hom$ is the group of homotheties acting on the Euclidean plane. Note that it does not inherit a quotient group structure from $\SL$ since $\SU$ is not a normal subgroup of $\SL$.} As we work in a cosmological context, adjoint-invariance is imposed in addition to right-invariance and simplicity and therefore, the final domain in group representation is given by
\begin{equation}\label{eq:final domain in group variables}
\Hom^3/\text{Ad}_{\SL}.
\end{equation}
Again, we apply the non-commutative Fourier transform to find
\begin{equation}
\Hom^3/\text{Ad}_{\SL}\overset{\text{nc FT}}{\longrightarrow}\mathfrak{D}\cong \hom^3/\text{Ad}_{\SL},
\end{equation}
where $\hom$ is the Lie algebra of $\Hom$ and where we have used again property \eqref{eq:adjoint on group is adoint on bivectors}, which translates the adjoint action of $\SL$ on group variables to the adjoint action of $\SL$ on bivectors. As a consequence of commuting imposition of constraints in the extended formalism, we would arrive at the domain
\begin{equation}
\mathfrak{D}\cong \hom^2,
\end{equation}
by first imposing right and adjoint-covariance\footnote{As shown explicitly in Section \ref{subsec:Spin Representation of Condensate Wavefunction}, right- and adjoint-covariance are equal to right-covariance and left-invariance. Under integration of the timelike normal, each of the conditions divide out one factor of $\SL^4$, resulting in the domain $\SL^2$.} and then simplicity. Clearly, the domain in this form is six-dimensional. We conclude that although the space $\hom^3/\text{Ad}_{\SL}$ appears to be three-dimensional,  the orbits of the adjoint $\SL$ action are really only three-dimensional and therefore, the domain is six-dimensional in total, corresponding to the six dimensions of minisuperspace.

Based on the studies in~\cite{Gielen:2013naa, Gielen:2014ila}, we make the diffeomorphism of $\mathfrak{D}$ and the configurations space of homogeneous 3-geometries explicit for the extended Lorentzian BC GFT model. First, note that $\mathfrak{D}$ describes three independent bivectors $B_i\in\spl$, $i\in\{1,2,3\}$ upon which simplicity is imposed, modulo any simultaneous Lorentz transformation. For every bivector, the linear simplicity constraint \eqref{eq:simplicity on bivectors} with respect to the timelike normal $X_A = (1,0,0,0)$ implies the vanishing of its electric part $B^{0a}$. Now, we relate the three bivectors $B_i^{AB}$ to three vectors in Minkowski space $e_i^A\in\R^{1,3}$ via~\cite{Freidel:1999rr}
\begin{equation}\label{eq:classical solution of constraint equation}
B_i^{AB} = \tensor{\varepsilon}{_i^{jk}}e_j^A e_k^B,
\end{equation}
where $e_i^A$ is the tetrad with Lorentz index $A\in\{0,1,2,3\}$ and spatial index $i$. Since $B^{0a} = 0$, the three Minkowski vectors need to satisfy $e_i^0 = 0$, which corresponds exactly to the time-gauge fixing that is crucial in the construction of LQG~\cite{Rovelli:2014ssa}. Notice, that the map defined by Eq.~\eqref{eq:classical solution of constraint equation} exhibits a $\mathbb{Z}_2$-symmetry, since $\{e_i^A\}$ and $\{-e_i^A\}$ define the same triple of bivectors. 

Now, the map from $\mathfrak{D}$ to the domain of minisuperspace, given by GL$(3)/$O$(3)$, is defined explicitly by~\cite{Gielen:2013naa}
\begin{equation}\label{eq:map to metric coefficients}
\begin{aligned}
\mathfrak{D} & \longrightarrow \text{GL}(3)/\text{O}(3)\\[7pt]
(B_1, B_2, B_3) & \longmapsto g_{ij} = e_{iA}e_j^A = \frac{1}{8\tr(B_1 B_2 B_3)}\tensor{\varepsilon}{_i^{kl}}\varepsilon_{jmn}\left(B_k^{AB}B^m_{AB}\right)\left(B_l^{CD}B^n_{CD}\right).
\end{aligned}
\end{equation}
Clearly, the map is well-defined with respect to the adjoint action of $\SL$ since all the Lorentz indices are contracted. It is also evident, that the map is only well-defined for triples of bivectors that satisfy $\tr(B_1 B_2 B_3)\neq 0$, which is an extra condition on the elements of $\mathfrak{D}$. However, this is exactly the three-dimensional non-degeneracy condition that needs to be imposed by hand, which we require at this point. 

On the other hand, the coefficients $g_{ij}$ determine three vectors $e_i^a$ up to a sign, which is however irrelevant if the vectors are further mapped to the bivectors $B_i$, because of the $\mathbb{Z}_2$-symmetry of Eq.~\eqref{eq:classical solution of constraint equation}. Hence, the map defined by Eq.~\eqref{eq:map to metric coefficients} is a diffeomorphism of $\mathfrak{D}$ and $\text{GL}(3)$/$\text{O}(3)$. A diagrammatic summary of this diffeomorphism is given in Figure \ref{fig: isomorphism of domain and mss}, where at every step, integration over the timelike normal is implicitly understood.

\begin{figure}[ht]
\centering
\begin{tikzcd}[row sep = 20pt, column sep = 0 pt]
\SL^4 \arrow[d, "\text{closure}"', "\text{Eq.}~\eqref{eq:covariance under right}"]
\\
\SL^3 \arrow[d, "\text{simplicity}"', "\text{Eq.}~\eqref{eq:generalized simplicity}"]
\\
\Hom^3 \arrow[d,"\text{adjoint invariance}"', "\text{Eq.}~\eqref{eq:adjoint covariance becomes adjoint invariance}"]
\\
\Hom^3/\text{Ad}_{\SL} \arrow[d, "\text{nc FT}"]
\\  
\hom^{\oplus 3}/\text{Ad}_{\SL} \arrow[d, "\text{map to $g_{ij}$}"', "\text{Eq.}~\eqref{eq:map to metric coefficients}"]
\\
\text{GL}(3)/\text{O}(3) & {}`
\end{tikzcd}
\caption{Diffeomorphism between the domain of the condensate wavefunction and minisuperspace of homogeneous spatial geometries.}
\label{fig: isomorphism of domain and mss}
\end{figure}

Let us elaborate on why the connection of domain and minisuperspace is conceptually important and serves as a self-consistency check of our cosmological interpretation (or, conversely, as a motivation for it).

First, the diffeomorphism should exist by the construction explained in Section \ref{sec:GFT Condensates}. Following~\cite{Gielen:2013naa}, a classical tetrahedron that is embedded at a point $x\in\Sigma$ in a spatial hypersurface $\Sigma$ defines three independent vectors $\vb{e}_i$ in the tangent space at $x$, which correspond to the three edges incident at the vertex of the tetrahedron. These three vectors are unique up to a simultaneous $\text{O}(3)$-rotation and define the six independent metric coefficients by
\begin{equation}
g_{ij}(x)
=
g(x)(\vb{e}_i,\vb{e}_j).
\end{equation}
From this perspective, the intrinsic geometry of an embedded tetrahedron carries exactly the metric information at this point.  Therefore, the domain $\mathfrak{D}$ must be diffeomorphic to $\textrm{GL}(3)/\textrm{O}(3)$~\cite{Gielen:2014ila}, only including the intrinsic geometric information. Note, that the dimension of minisuperspace being six, corresponds to the six degrees of freedom of a classical tetrahedron, parametrized for example by its six edge lengths or four areas and two dihedral angles~\cite{Pereira2010}. 

Second, the diffeomorphism of domain and minisuperspace is an important link between the kinematics of the GFT condensate cosmology ansatz, the configuration space of classical spatially homogeneous cosmologies~\cite{Ellis2012}, Wheeler-de Witt quantum cosmology~\cite{Kiefer2012,Hamber2009} and Loop Quantum Cosmology (LQC)~\cite{Bojowald2008,Ashtekar:2021kfp,Banerjee:2011qu}. Such a connection has been already established for GFT condensates based on EPRL-like models~\cite{Oriti:2016qtz,Gielen:2014ila,Gielen:2013naa}.

After these clarifications, we would like to point out similarities to the condensate approach based on EPRL-like models~\cite{Oriti:2016qtz, deCesare:2017ynn, Pithis:2019tvp, Gielen:2014ila, Oriti:2016acw} where the domain of the group field is $\SU\setminus\SU^4/\SU$, so four copies of $\SU$ with diagonal left- and right-invariance. The right-invariance accounts for the closure condition, meaning that the four triangles of a GFT quantum close to form a tetrahedron. In~\cite{Oriti:2016qtz} it is stated that the left-invariance accounts for the independence of a specific embedding of tetrahedra in a spatial hypersurface. In the case of $\SU\setminus\SU^4/\SU$ there is an obvious isomorphism
\begin{equation}
\SU\setminus\SU^4/\SU
\cong
 \left(\SU^4/\SU\right)/\text{Ad}_{\SU}.
\end{equation} 
Again, the interpretation applies that the adjoint action induces rotations of bivectors. Therefore, the domain only encodes the physical degrees of freedom, which is also shown in the diffeomorphism to minisuperspace. Changing variables via the non-commutative Fourier transform on $\SU$~\cite{Guedes:2013vi,Oriti:2018bwr}, the domain is given by
\begin{equation}
\mathfrak{D}_{\text{EPRL-GFT}}
\cong
\su^3/\text{Ad}_{\SU},
\end{equation}
which is diffeomorphic to minisuperspace by the same mechanism as shown above for the BC model~\cite{Gielen:2014ila}.

With the domain of the reduced condensate wavefunction now specified, the next goal is to derive dynamical equations arising from the equations of motion of $\sigma$. 

As it turns out, performing actual calculations and imposing isotropy on solutions is most convenient in spin representation. Therefore, the spin representation of the reduced condensate wavefunction and of the semi-classical action are given in Sections \ref{subsec:Spin Representation of Condensate Wavefunction} and \ref{subsec:Spin Representation of the Action and Regularization}, based on which the cosmological dynamics are then derived in Section \ref{sec:Evolution of Effective Cosmology}.

\subsection{Spin representation of reduced condensate wavefunction}\label{subsec:Spin Representation of Condensate Wavefunction}

In the spin representation, the isotropic restriction on solutions is implemented rather straightforwardly and the equations of motion simplify drastically. Partly, the expansion has been already achieved in Section \ref{subsec:Spin Representation and Regularization of Group Field} for the GFT field. However, specific to the reduced condensate wavefunction, we introduced the third condition of adjoint-covariance. As this additional covariance will lead to a divergence of the action, we are going to perform a second regularization, the details of which are given in Appendix \ref{appendix:Recoupling Theory of SL2C}.

For computational ease, instead of adjoint-covariance we impose a left-invariance. The equivalence between these conditions is seen by the following consideration. Let $\sigma(g_i;X)$ transform covariantly under right and adjoint action, i.e.
\begin{equation}
\sigma(g_v;X;\phi)
=
\sigma(g_v h^{-1}; h\cdot X;\phi)
=
\sigma(kg_v k^{-1}; k\cdot X;\phi),\quad \forall h,k\in\SL.
\end{equation}
Using both covariances combined, we find
\begin{equation}
\sigma(g_v;X;\phi)
=
\sigma(kg_v h^{-1} k^{-1}; kh\cdot X;\phi)
\overset{\tilde{h}
\defeq kh}{=}\sigma(kg_v\tilde{h}^{-1}; \tilde{h}\cdot X;\phi )
\overset{\text{right cov.}}{=}
\sigma(k g_v; X;\phi),
\end{equation}
which means that a right- and adjoint-covariant function is effectively given by a right-covariant and left-invariant function. 

Imposing left-invariance requires recoupling theory of $\SL$ since integrals of the form
\begin{equation}
\int\limits_{\SL}\dd{h}\prod_{i=1}^4 D^{(\rho_i,0)}_{j_i m_i l_i n_i}(h)
\end{equation}
appear. Tools from recoupling theory of $\SL$ that are required in this work are given in the appendix based on~\cite{Speziale:2016axj}. Applying Equation \eqref{eq:simplified integral over four SL2C matrices}, which is a notationally simplified version of Eq.~\eqref{eq:Integral over four SL2C matrices}, the expansion of the reduced condensate wavefunction in terms of representation variables is given by
\begin{equation}\label{eq:wavefunction with right covariance, simplicity, left invariance}
\begin{aligned}
\sigma(g_v;X;\phi)
& =
\left[\prod_i \int\dd{\rho_i}4\rho_i^2\sum_{j_i m_i}\right]\int\dd{\rho_{12}}\sum_{\nu_{12}}4\left(\rho_{12}^2+\nu_{12}^4\right)\times\\[7pt]
& \times
\sigma^{\rho_i,(\rho_{12},\nu_{12})}(\phi)\overline{\mqty((\rho_i,\nu_i) \\ j_i m_i)^{(\rho_{12},\nu_{12})}} \prod_{i}D^{(\rho_i,0)}_{j_i m_i 00 }(g_ia).
\end{aligned}
\end{equation}
Finally, we consider the expansion of the reduced condensate wavefunction under integration over the timelike normal
\begin{equation}\label{eq:wavefunction with all invariances and integration over normal}
\begin{aligned}
\int\limits_{\HH}\dd{X}\sigma(g_v;X;\phi) 
&=
\left[\prod_i \int\dd{\rho_i}4\rho_i^2\sum_{j_i m_i l_i n_i}\right]\int\dd{\rho_{12}}\sum_{\nu_{12}}4\left(\rho_{12}^2+\nu_{12}^4\right)\times\\[7pt]
& \times
\sigma^{\rho_i,(\rho_{12},\nu_{12})}(\phi)\overline{\mqty((\rho_i,\nu_i) \\ j_i m_i)^{(\rho_{12},\nu_{12})}}B^{\rho_i}_{l_i n_i} \prod_{i}D^{(\rho_i,0)}_{j_i m_i l_i n_i }(g_i).
\end{aligned}
\end{equation}

Applying expansion~\eqref{eq:wavefunction with right covariance, simplicity, left invariance} to the kinetic kernel and expansion~\eqref{eq:wavefunction with all invariances and integration over normal} to the vertex kernel, we derive the action and its regulated form in the following section.

\subsection{Effective action in spin representation and regularization}\label{subsec:Spin Representation of the Action and Regularization}

Forming the expectation value of the operator valued action \eqref{eq:divergent BO action} with respect to the coherent peaked state $\ket{\sigma_{\epsilon};\phi_0,\pi_0}$ yields the semi-classical action for the condensate wave function $S[\sigma_{\epsilon},\bar{\sigma}_{\epsilon}]$
\begin{equation}
S[\sigma_{\epsilon},\bar{\sigma}_{\epsilon}] \defeq \expval{S[\hat{\varphi},\hat{\varphi}^{\dagger}]}{\sigma_{\epsilon};\phi_0,\pi_0}.
\end{equation}
Adapting the procedure of~\cite{Marchetti:2020umh} to the extended Lorentzian BC GFT model, the reduced condensate wavefunction is expanded around the peaking value $\phi_0$ yielding a simplification of the semi-classical action. The resulting effective action $S[\sigma,\bar{\sigma}]$ governs the relational evolution of the reduced condensate wavefunction $\sigma(g_v;X;\phi)$, the spin representation of which we specify in the following.

For the kinetic term, we make use of Eq.~\eqref{eq:expansion of kinetic kernel} and expand the kinetic kernel and the reduced condensate wavefunction around the peaking value $\phi_0$ and truncate at order $\epsilon$~\cite{Marchetti:2020umh}. As a result of these approximations, the effective kinetic action is given by
\begin{equation}\label{eq:kinetic action in spin representation}
\begin{aligned}
K_{\text{eff}}
& =
\int\dd{\phi_0}\left[\prod_i \int\dd{\rho_i}4\rho_i^2\right]\int\dd{\rho_{12}}\sum_{\nu_{12}}4\left(\rho_{12}^2+\nu_{12}^2\right)\bar{\sigma}^{\rho_i,(\rho_{12},\nu_{12})}(\phi_0)\bigg{(}\mathscr{K}_{(2)}^{\rho_i,(\rho_{12},\nu_{12})}+\\[7pt]
&-
\frac{2\mathscr{K}_{(0)}^{\rho_i,(\rho_{12},\nu_{12})}}{\epsilon(\epsilon\pi_0^2-1)} -2i\tilde{\pi}_0\mathscr{K}_{(0)}^{\rho_i,(\rho_{12},\nu_{12})}\pdv{}{\phi_0} +{}\mathscr{K}_{(0)}^{\rho_i,(\rho_{12},\nu_{12})}\pdv[2]{}{\phi_0}\bigg{)}\sigma^{\rho_i,(\rho_{12},\nu_{12})}(\phi_0),
\end{aligned}
\end{equation}
where
\begin{equation}
\tilde{\pi}_0\defeq\frac{\pi_0}{\epsilon\pi_0^2-1}.
\end{equation}
Considering the two lowest-order coefficients of Eq.~\eqref{eq:expansion of kinetic kernel} in spin representation, the residual coefficients $\mathscr{K}_{(i)}$ are  defined by
\begin{equation}
\mathcal{K}^{\rho_i\rho_i',(\rho_{12},\nu_{12})(\rho_{12}',\nu_{12}')}_{(i)}
=
\mathscr{K}_{(i)}^{\rho_i,(\rho_{12},\nu_{12})}\frac{\delta(\rho_{12}-\rho_{12}')\delta_{\nu_{12},\nu_{12}'}\prod_i \delta(\rho_i-\rho_i')}{4\left(\rho_{12}^2+\nu_{12}^2\right)4^4 \prod_i\rho_i^2},
\end{equation}
and contain any residual freedom that entered with the coupling to the scalar field and possibly from renormalization group arguments, such as a Laplace operator~\cite{BenGeloun:2013mgx}. Notice, that Eq.~\eqref{eq:kinetic action in spin representation} is already given in its regulated form, where the divergence arose from summing over $\SL$-intertwiners, see Eq.~\eqref{eq:divergence of SL2C intertwiners}.

Similarly, the effective vertex action is given by 
\begin{equation}\label{eq:vertex action in spin representation}
\begin{aligned}
V_{\text{eff}} &= \frac{\lambda}{5}\frac{\mathcal{N}_{\epsilon}^3}{\sqrt{2\pi\epsilon}}\e^{\pi_0^2\epsilon/2}\int\dd{\phi_0}\left[\prod_{a=1}^{10}\int\dd{\rho_a}4\rho_{a}^2\sum_{j_a m_a}\prod_{b=11}^{15}\int\dd{\rho_b}\sum_{\nu_b}4\left(\rho_b^2+\nu_b^2\right)\right]\overline{\{15\rho\}}\{10\rho\}_{\text{BC}}\times\\[7pt]
& \times
\sigma^{\rho_1 \rho_2 \rho_3 \rho_4,(\rho_{11},\nu_{11})}(\phi_0)\sigma^{\rho_4 \rho_5 \rho_6 \rho_7,(\rho_{12},\nu_{12})}(\phi_0)\sigma^{\rho_7 \rho_3 \rho_8 \rho_9,(\rho_{13},\nu_{13})}(\phi_0)\times\\[7pt]
& \times
\sigma^{\rho_9 \rho_6 \rho_2 \rho_{10},(\rho_{14},\nu_{14})}(\phi_0)\sigma^{\rho_{10} \rho_8 \rho_5 \rho_1 ,(\rho_{15},\nu_{15})}(\phi_0)+c.c.,
\end{aligned}
\end{equation}
where we recognize the regulated Barrett-Crane $\{10\rho\}_{\text{BC}}$-symbol defined in Eq.~\eqref{eq:definition of regulated BC 10p}. Due to effective left-invariance, an $\SL$-$\{15\rho\}$-symbol defined in Eq.~\eqref{eq:definition of SL2C 15p symbol} arises, which is the contraction of five $\SL$-intertwiners. The regularization of the $\{15\rho\}$-symbol works analogous to the $\{10\rho\}$-symbol, as described in Section \ref{subsec:Spin Representation and Regularization of Group Field}. Again, we only work with the regularized action in the following and drop any indication of regularization for notational ease. 

In contrast to EPRL-like GFT models~\cite{Oriti:2016qtz,Gielen:2014ila,Gielen:2013naa,Pithis:2019tvp,Pithis:2016cxg} in the cosmological  application, where the vertex kernel (so far only) implicitly encodes the relation of $\SU$- and $\SL$-data, the vertex kernel of the cosmological BC model is determined without any residual freedom and the use of the full Lorentz group is made explicit. We emphasize that the spectra of geometric operators are discrete in the EPRL-like approach and continuous in the BC approach.  Furthermore, the Barbero-Immirzi parameter $\gamma$ enters geometric spectra and the simplicity constraint in the EPRL-like models while it is absent in the BC model. Hence, the quantum geometries described by both models are quite different. We will now see whether these differences at the level of the microscopic description of quantum geometry lead to a different effective continuum cosmological dynamics. 

\section{Effective cosmological evolution}\label{sec:Evolution of Effective Cosmology}

 After specifying an appropriate microscopic quantum analogue of isotropy in the following section, we derive the volume dynamics in Section~\ref{subsec:Effective Equations and Volume Dynamics}, study the classical limit in Section~\ref{subsec:Classical Limit} and shed light on the bounce mechanism in GFT condensate cosmology in Section~\ref{subsec:Emergent Quantum Bounce}. Subsequently, we study single-label condensates in Section~\ref{subsec:Single-Label Condensates, Quantum Corrected Friedmann Equations and LQC} and list some phenomenological implications of the BC GFT condensate cosmology in section~\ref{subsec:Phenomenological Implications}.

Given by the sum of Eq.~\eqref{eq:kinetic action in spin representation} and Eq.~\eqref{eq:vertex action in spin representation}, the effective action of the reduced condensate wavefunction yields the effective equations of motion via
\begin{equation}
\fdv{S_{\text{eff}}[\sigma,\bar{\sigma}]}{\bar{\sigma}^{\rho_i,(\rho_{12},\nu_{12})}(\phi_0)} = 0,
\end{equation}
using the derivative
\begin{equation}
\fdv{\bar{\sigma}^{\rho_1 \rho_2 \rho_3 \rho_4,(\rho_{12},\nu_{12})}(\phi_0)}{\bar{\sigma}^{\rho_1' \rho_2' \rho_3' \rho_4',(\rho_{12}',\nu_{12}')}(\phi'_0)}
=
\delta(\phi_0-\phi'_0)\delta(\rho_{12}-\rho_{12}')\delta_{\nu_{12},\nu_{12}'}\prod_{i=1}^4\delta(\rho_i-\rho_i').
\end{equation}
We note that these equations of motion for the condensate field assume the shape of non-linear tensor equations which are very difficult to solve in full generality, see for instance~\cite{Fairbairn:2007sv,Girelli:2009yz,Girelli:2010ct,Livine:2011yb,BenGeloun:2018eoe}. Nevertheless, we are going to show in a moment how an emergent cosmological evolution in terms of the relational clock $\phi$ can arise, the phenomenological implications of which we study extensively thereafter. 

We remind that homogeneity is obtained in a coarse grained sense by having reduced the full dynamics to an effective equation for the mean field, whose domain matches minisuperspace. To recover the simplest form of cosmological dynamics, we now have to restrict to isotropic data.

\subsection{Isotropic restriction}\label{subsec:Isotropic Restriction and Fixing of Left Intertwiner}

In accordance with~\cite{Oriti:2016qtz,Gielen:2016dss,Pithis:2016cxg,Pithis:2019tvp}, a possible\footnote{The restriction to states that yield a isotropic spatial geometry in the continuum is not unique. A different notion of isotropy, given by tri-rectangular tetrahedra (which translates to three of the four bivectors being orthogonal), has been studied in~\cite{Pithis:2019tvp,Pithis:2016cxg}, which led however to the same condensate dynamics.} restriction to a state that yields an isotropic continuum spacetime is given by considering condensate wavefunctions corresponding to fundamental tetrahedra of equal faces area. 
 
In contrast with the EPRL-like models (and the canonical LQG quantum geometry), the area operator acting on simplicial triangles of a Barrett-Crane tetrahedron is given by 
\begin{equation}
\hat{A}_{\text{BC}}
=
\sqrt{B_{AB}B^{AB}}
=
\sqrt{\cas_2(\SL)},
\end{equation}
in terms of the first Casimir of $\SL$\footnote{In the extended formalism, boundary states are given by so-called projected spin networks~\cite{Livine:2002ak,Alexandrov:2002br} for which the eigenvalues $A_{\text{psn}}$ of the area operator are given by~\cite{Alexandrov:2001pa} $\label{eq:proportionality PSN area}
A_{\text{psn}}\propto \sqrt{j(j+1)+\rho^2-\nu^2+1}$, where the index $j$ is associated to a pair of edge and vertex~\cite{Livine:2002ak}. Since simplicity is imposed on the edges where they meet at a vertex, $\nu$ and also $(j,m)$ are set to zero. Thus, for a simplicial tetrahedron as defined by the domain \eqref{eq:final domain in group variables}, considered as projected spin network, the area $A_{\text{psn}}$ reduces to the ordinary area $A_{\text{BC}}$, given in Eq.~\eqref{eq:proportionality BC area}.} defined in Eq.~\eqref{eq:definition of cas1}, with eigenvalues
\begin{equation}\label{eq:proportionality BC area}
\hat{A}_{BC}\ket{(\rho,0);00} =  \sqrt{\rho^2+1}\ket{(\rho,0);00},
\end{equation}
where $\ket{(\rho,0);jm}$ is the canonical basis of the $(\rho,0)$ representation space of $\SL$. 
It is important to note, that in addition to different representation labels, the area spectra of the BC model differ from the ones in the EPRL model by the absence of the Barbero-Immirzi parameter $\gamma$. Furthermore, we remark that the spectrum of $\hat{A}_{\text{BC}}$ is non-zero for any value of $\rho\in\R$, while the area spectrum in EPRL or LQG is zero for $j=0$ (when using the symmetric quantization map). This is often used as a motivation for imposing, in the kinematical space of LQG, cylindrical consistency conditions effectively removing these configurations from the theory~\cite{Thiemann2007a}. Such cylindrical consistency is absent, however, in the EPRL model.

If $i\in\{1,2,3,4\}$ denotes the $i$-th triangle with its area fully determined by $\rho_i$ as given in Equation \eqref{eq:proportionality BC area}, the condition of equal area translates to the equivalence of all the $\rho_i$
\begin{equation}
\rho_i = \rho,\quad\forall i\in\{1,2,3,4\}.
\end{equation}
Equating areas of triangles is not enough for a fully isotropic configuration of tetrahedra and, as a consequence, of homogeneous universe geometries.

In addition, we need to consider the intertwiner labels $(\rho_{12},\nu_{12})$, which play a similar role to the left $\SU$-intertwiner $\iota_l$, appearing in~\cite{Oriti:2016qtz,deCesare:2017ynn}. As the following arguments ~\cite{Baez:1999tk,Oriti:2003wf} suggest, it is reasonable to fix $(\rho_{12},\nu_{12})$ to some value $(\rho_*,\nu_*)$.

For generic classical tetrahedra, the space of geometries is six-dimensional, parametrized for instance by four triangle areas and two areas of parallelograms which have vertices in the middle of edges.  This is also reflected by the diffeomorphism of domain and minisuperspace as derived in Section~\ref{subsec:Isomorphism of Domain and Minisuperspace}. However at the quantum level, the degrees of freedom are reduced, as indicated by the non-commutativity of parallelogram area operators. Indeed, while all the triangle area operators commute among each other and with each of the parallelogram area operators, the latter do not commute among themselves. Four area operators and one of the parallelogram area operators form a set of maximally commuting operators and therefore, determining the areas and one parallelogram area of the quantum tetrahedron leaves the other parallelogram area operator fully randomized by the uncertainty principle. While this applies to generic quantum tetrahedra, the BC quantum tetrahedron represents a special case where the fifth parameter is not independent of the other four and the quantum tetrahedron is fully characterized by $\rho_1,...,\rho_4$. Also for this reason, the BC intertwiner defined in Eq.~\eqref{eq:definition of BC intertwiner} is uniquely determined by the incident Lorentz representations, in contrast to e.g. $\SU$-intertwiners~\cite{Rovelli:2014ssa}. 
Following the tentative reasoning in \cite{Oriti:2016qtz}, thus interpreting the second intertwiner appearing in our condensate wavefunction as the parallel transported one of the BC intertwiner, thus not carrying new dynamical degrees of freedom, we restrict to states where $(\rho_{12},\nu_{12}) = (\rho_*,\nu_*)$ is fixed.

Hence, we restrict our attention to specific solutions of the form
\begin{equation}
\sigma^{\rho_1 \rho_2 \rho_3 \rho,(\rho_{12},\nu_{12})}(\phi)
=
\sigma_{\rho}(\phi) \delta_{\nu_{12},\nu_*}\delta(\rho_{12}-\rho_*)\prod_{i=1}^3\delta(\rho_i-\rho).
\end{equation}
Evaluating the effective action on solutions that satisfy these conditions, we find
\begin{equation}\label{eq:action after isotropic restriction}
\begin{aligned}
S_{\text{eff}}[\sigma,\bar{\sigma}]
& =
\int\dd{\phi_0}\int\dd{\rho}\bigg{[} A_{\rho}\abs{\partial_{\phi_0}\sigma_{\rho}(\phi_0)}^2+2i\tilde{\pi}_0A_{\rho}\bar{\sigma}_{\rho}(\phi_0)\partial_{\phi_0}\sigma_{\rho}(\phi_0)+\\[7pt]
& +
\left(\frac{2A_{\rho}}{\epsilon(\epsilon\pi_0^2-1)}+B_{\rho}\right)\abs{\sigma_{\rho}(\phi_0)}^2-\frac{1}{5}u_{\rho}\sigma_{\rho}(\phi_0)^5-\frac{1}{5}\bar{u}_{\rho}\bar{\sigma}_{\rho}(\phi_0)^5 \bigg{]},
\end{aligned}
\end{equation}
with the shorthand notation 
\begin{align}
& A_{\rho} \defeq 4^4\rho^8 4\left(\rho_*^2+\nu_*^2\right)\;\mathscr{K}_{(0)}^{\rho\rho\rho\rho,(\rho_*,\nu_*)}\label{eq:definition of A},\\[7pt]
& B_{\rho} \defeq -4^4 \rho^8 4\left(\rho_*^2+\nu_*^2\right)\;\mathscr{K}_{(2)}^{\rho\rho\rho\rho,(\rho_*,\nu_*)},\label{eq:definition of B}\\[7pt]
& u_{\rho} \defeq \frac{\mathcal{N}_{\epsilon}^3}{\sqrt{2\pi\epsilon}}w_{\rho},\label{eq:definition of u}\\[7pt]
& w_{\rho} \defeq \lambda\; 4^{10}\rho^{20} 4^5\left(\rho_*^2+\nu_*^2\right)^5\overline{\{15\rho\}}\bigg{\vert}_{\substack{\rho_a = \rho,\forall a\in\{1,...,10\} \\ \rho_b = \rho_*, \forall b\in\{11,...,15\}}}\{10\rho\}_{\text{BC}}\bigg{\vert}_{\rho_a = \rho,\forall a\in\{1,...,10\}},\label{eq:definition of w}
\end{align}
wherein the factors of $4$, $\rho^2$ and $(\rho_*^2+\nu_*^2)$ stem from the Plancherel measure on the space of functions on $\SL$~\cite{Speziale:2016axj}.\footnote{Notice that the action~\eqref{eq:action after isotropic restriction} is real despite the imaginary term proportional to $2i$. Forming the complex conjugate of this term and performing a partial integration shows that it is indeed real.}\textsuperscript{,}\footnote{Notice that at this point the relation to a (modified) FLRW cosmology is not yet apparent. This is established in Secs.~\ref{subsec:Effective Equations and Volume Dynamics} and~\ref{subsec:Classical Limit} by studying the dynamics of the expectation value of the volume operator. However, should the action~\eqref{eq:action after isotropic restriction} exhibit a hidden conformal symmetry, it could very well be that the connection to FLRW cosmology could already be established at this point. This would follow along the lines of recent results on mapping of FLRW cosmology onto conformal mechanics, see~\cite{BenAchour:2019ufa,BenAchour:2020njq,Achour:2021lqq}.} As a consequence of the isotropic restriction and the intertwiner fixing, we have obtained a decoupling of the reduced condensate wavefunctions in the interaction term, that is, we obtain local interactions of the form $\propto w_{\rho}\sigma_{\rho}(\phi)^5$ instead of non-local interactions similar to $\propto w_{\rho_1 \rho_2 \rho_3 \rho_4 \rho_5}\sigma_{\rho_1}(\phi)\sigma_{\rho_2}(\phi)\sigma_{\rho_3}(\phi)\sigma_{\rho_4}(\phi)\sigma_{\rho_5}(\phi)$. The only difference  to the action of isotropic condensate wavefunctions in the EPRL-like model in~\cite{Marchetti:2020umh} lies in the definition of the coefficients $A_{\rho},B_{\rho},w_{\rho}$ and in the fact that the remaining representation label $\rho$ is continuous instead of discrete. 

This may turn out to be important differences for any dynamical regime or computation of observables for which the precise functional form of the model coefficients is relevant. However, as we will see this is not the case for the main qualitative features of the emergent homogeneous and isotropic cosmology.

The subsequent analysis of the effective equations of motion and their phenomenological implications, including also the appearance of the quantum bounce, is done analogously to~\cite{Oriti:2016qtz} and~\cite{Marchetti:2020umh}.

In the following section, we derive the effective equations of motion from the action \eqref{eq:action after isotropic restriction}, determine the conserved quantities it defines and ultimately obtain the emergent Friedmann equations modified by quantum corrections, implying the resolution of the initial singularity into a quantum bounce.

\subsection{Effective cosmological equations and volume Dynamics}\label{subsec:Effective Equations and Volume Dynamics}

In order to extract the relational evolution of the BC condensate, we derive the effective equations of motion of the reduced condensate wavefunction restricted to isotropy by varying the effective action in \eqref{eq:action after isotropic restriction}, leading to
\begin{equation}\label{eq:nonlineom}
\sigma''_{\rho}(\phi_0) -2i\tilde{\pi}_0\sigma'_{\rho}(\phi_0)-M_{\rho}^2\sigma_{\rho}(\phi_0)+\bar{u}_{\rho}\bar{\sigma}^4_{\rho}(\phi_0) = 0,
\end{equation}
where 
\begin{equation}
M_{\rho}^2
\defeq
\frac{2}{\epsilon(\epsilon\pi_0^2-1)}+\frac{B_{\rho}}{A_{\rho}},
\end{equation}
and prime denotes differentiation with respect to $\phi_0$.\footnote{Given the similarities of the equation of motion of the condensate field~\eqref{eq:nonlineom} to the non-linear Schr{\"o}dinger or Klein-Gordon equations~\cite{Vachaspati:2006zz}, it might likewise admit solitary wave solutions. We leave the interesting question of the integrability of this non-linear differential equation and the quantum cosmological interpretation thereof to future research. Notice, however, that numerical solutions in the case of the EPRL-like models have been investigated in Refs.~\cite{deCesare:2016rsf,Pithis:2016cxg} which, given our results, can be carried over to the case of the BC model straightforwardly.} In addition, the effective action exhibits a conserved quantity
\begin{equation}
\tilde{\mathcal{E}}_{\rho} 
=
\abs{\sigma'_{\rho}}^2-M_{\rho}^2\abs{\sigma_{\rho}}^2+\frac{2}{5}\mathfrak{Re}\left[u_{\rho}\sigma_{\rho}^5\right],
\end{equation}
arising from translational invariance of the integrand of \eqref{eq:action after isotropic restriction} with respect to $\phi_0$.

As we have elaborated in Section \ref{subsec:Coherent Peaked Condensate States}, the coherent peaked condensate states should be expected to fail to approximate solutions of the quantum equations of motion for strong interactions. This happens if either the coupling constants in front of interactions or the particle number (thus, the field density) are too large. We therefore focus our attention to a regime of negligible interactions.\footnote{Notice, that both $\frac{\mathcal{N_{\epsilon}}^3}{\sqrt{2\pi\epsilon}}$ and $\e^{\epsilon\pi_0^2/2}$, entering the definition of $u_{\rho}$ in Eq.~\eqref{eq:definition of u}, are large and therefore, the requirement of negligible interactions poses conditions on microscopic couplings $A_{\rho}$ and $w_{\rho}$~\cite{Marchetti:2020umh}.}

The equations of motion simplify to 
\begin{equation}\label{eq:approximate eom}
\sigma''_{\rho}(\phi_0) -2i\tilde{\pi}_0\sigma'_{\rho}(\phi_0)-M_{\rho}^2\sigma_{\rho}(\phi_0) \approx 0,
\end{equation}
and $\tilde{\mathcal{E}}_{\rho}$ takes the form
\begin{equation}\label{eq:approximate energy}
\tilde{\mathcal{E}}_{\rho} 
\approx
\abs{\sigma'_{\rho}}^2-M_{\rho}^2\abs{\sigma_{\rho}}^2.
\end{equation}
In addition to $\tilde{\mathcal{E}}_{\rho}$ a second conserved quantity $\mathcal{Q}_{\rho}$ arises under the assumption of negligible interactions, which is associated to a $\text{U}(1)$-symmetry
\begin{equation}
\sigma_{\rho}(\phi)\longmapsto \e^{i\alpha}\sigma_{\rho}(\phi),\quad\alpha\in\R.
\end{equation}
From Noether's theorem, the expression of $Q_{\rho}$ follows
\begin{equation}\label{eq:definition of Q}
\mathcal{Q}_{\rho}
\defeq
-\frac{i}{2}
\left[\bar{\sigma}_{\rho}\sigma'_{\rho} - \bar{\sigma}'_{\rho}\sigma_{\rho}-2i\tilde{\pi}_0\abs{\sigma_{\rho}}^2\right].
\end{equation}
As it will turn out, defining a third conserved quantity
\begin{equation}\label{eq:definition of E}
\mathcal{E}_{\rho}
\defeq
\tilde{\mathcal{E}}_{\rho} - 2\mathcal{Q}_{\rho}\tilde{\pi}_0,
\end{equation}
leads to a simplification of the relational volume evolution.

In the next step, we introduce a split of the condensate wavefunction $\sigma_{\rho} = r_{\rho}\e^{i\theta_{\rho}}$ into radial and angular parts, splitting up Eq.~\eqref{eq:approximate eom} into two equations
\begin{align}
0 &= \left((\theta'_{\rho}-\tilde{\pi}_0)r_{\rho}^2\right)',\label{eq:conservation of Q}\\
0 &= r_{\rho}''-\frac{\mathcal{Q}_{\rho}^2}{r_{\rho}^3}-\mu_{\rho}^2r_{\rho}\label{eq:eom for r},
\end{align}
where we have defined 
\begin{equation}
\mu_{\rho}^2
\defeq
M_{\rho}^2-\tilde{\pi}_0^2.
\end{equation}
Eq.~\eqref{eq:conservation of Q} exactly corresponds to the conservation of the charge $Q_{\rho}$. On the other hand, the second equation contains the dynamical information of the (modulus of the) reduced condensate wavefunction which can be re-expressed using the definition of $\mathcal{E}_{\rho}$ in Eq.~\eqref{eq:definition of E}
\begin{equation}\label{eq:eom for r with E}
\mathcal{E}_{\rho}
=
\left(r_{\rho}'\right)^2+\frac{\mathcal{Q}_{\rho}^2}{r_{\rho}^2}-\mu_{\rho}^2r_{\rho}^2.
\end{equation}
Although having the form of an energy, we do not interpret $\mathcal{E}_{\rho}$ as a form of physical energy and instead keep considering it as an abstract conserved quantity. Notice that Eq.~\eqref{eq:eom for r} assumes the form of the equation of motion of a classical point particle in the potential 
\begin{equation}
    U_{\rho}(r_{\rho})=-\mu_{\rho}^2 r_{\rho}^2+\frac{\mathcal{Q}_{\rho}^2}{ r_{\rho}^2}.
\end{equation}

For a proper cosmological interpretation, we need to recast this equation into one for a geometric observable, i.e. as a relational evolution of the universe volume. 

Considering the expectation value of the volume operator defined in Eq.~\eqref{eq:definition of volume operator} with respect to the coherent peaked state yields
\begin{equation}
\langle\hat{V}\rangle_{\epsilon,\phi_o,\pi_0}
\defeq
\expval{\hat{V}}{\sigma_{\epsilon},\phi_0,\pi_0}
=
\int\dd{\phi}\int\dd{\rho}V_{\rho}\abs{\tilde{\sigma}_{\rho}(\phi)}^2\abs{\eta_{\epsilon}(\phi-\phi_0,\pi_0)}^2.
\end{equation}
After a lowest-order saddle-point approximation, which holds for a good clock with $\epsilon\ll 1$, the expectation value of the volume reduces to~\cite{Marchetti:2020umh}
\begin{equation}\label{eq:volume expectation value}
\langle\hat{V}\rangle_{\epsilon,\phi_0,\pi_0}
\approx
\int\dd{\rho}V_{\rho}\abs{\tilde{\sigma}_{\rho}(\phi_0)}^2
=
\int\dd{\rho}V_{\rho}r_{\rho}^2(\phi_0)
\equiv
V(\phi_0).
\end{equation}
Then, by deriving Eq.~\eqref{eq:volume expectation value} and using Eq.~\eqref{eq:eom for r with E}, we derive the following two volume evolution equations
\begin{align}
\left(\frac{V'}{3V}\right)^2
& =
\left(\frac{2\int\dd{\rho}V_{\rho}r_{\rho}\;\text{sgn}(r_{\rho}')\sqrt{\mathcal{E}_{\rho}-\frac{\mathcal{Q}_{\rho}^2}{r_{\rho}^2}+\mu_{\rho}^2r_{\rho}^2}}{3\int\dd{\rho}V_{\rho}r_{\rho}^2}\right)^2,\label{eq:effective CPS Friedmann equation 1}\\[7pt]
\frac{V''}{V}
& =
\frac{2\int\dd{\rho}V_{\rho}\left[\mathcal{E}_{\rho}+2\mu_{\rho}^2r_{\rho}^2\right]}{\int\dd{\rho}V_{\rho}r_{\rho}^2}.\label{eq:effective CPS Friedmann equation 2}
\end{align} 

In the following section, we define a classical limit and show under which conditions Eqs.~\eqref{eq:effective CPS Friedmann equation 1} and~\eqref{eq:effective CPS Friedmann equation 2} reduce to Friedmann equations of classical cosmology. Subsequently in Section~\ref{subsec:Emergent Quantum Bounce}, we analyze in-depth the quantum bounce emerging from the BC GFT condensate. Thereupon, we clarify the connection of the BC GFT condensate to LQC in Section~\ref{subsec:Single-Label Condensates, Quantum Corrected Friedmann Equations and LQC} and list several phenomenological implications of the BC GFT condensate approach in Section~\ref{subsec:Phenomenological Implications}.

\subsection{Classical Limit}\label{subsec:Classical Limit}

 Although the semi-classical limit is strictly speaking obtained for small curvature~\cite{Oriti:2016qtz,Gielen:2016dss,Ashtekar:2021kfp}, the contributions of large volume dominate in a regime of small curvature~\cite{Oriti:2016qtz} and therefore, we can simply consider the limit of large total volume. As the volume scales quadratically with $r_{\rho}^2$, taking the classical limit amounts to
\begin{equation}\label{eq:conditions of classical limit}
r_{\rho}^2 \gg  \abs{\mathcal{E}_{\rho}}/\mu_{\rho}^2,\quad r_{\rho}^4 \gg \mathcal{Q}_{\rho}^2/\mu_{\rho}^2,
\end{equation}
which leads to the effective Friedmann equations. 
\begin{align}
\left(\frac{V'}{V}\right)^2 & \longrightarrow \left(\frac{2\int\dd{\rho}V_{\rho}r_{\rho}^2\mu_{\rho}\text{sgn}(r_{\rho}')}{3\int\dd{\rho}V_{\rho}r_{\rho}^2}\right)^2,\\[7pt]
\frac{V''}{V} & \longrightarrow \frac{4\int\dd{\rho}V_{\rho}\mu_{\rho}^2r_{\rho}^2}{\int\dd{\rho}V_{\rho}r_{\rho}^2}.
\end{align}
Notice, that although $r_{\rho}^2$ is taken to be large compared to $\abs{\mathcal{E}_{\rho}}/\mu_{\rho}^2$ and $\mathcal{Q}_{\rho}^2/\mu_{\rho}^2$, it must remain not too large for the GFT interactions to remain negligible.

Assuming that there exist at least a dominant mode satisfying
\begin{equation}\label{eq:emergent G}
\mu_{\rho}^2 = 3\pi G,\quad \forall \rho\in\R,
\end{equation}
which represents a condition on the kinetic kernels ${}^{(0)}\mathscr{K}$, ${}^{(2)}\mathscr{K}$ and the clock parameters $\epsilon$ and $\pi_0$, we obtain the flat Friedmann equations in terms of the relational clock $\phi_0$\footnote{In the Hamiltonian formulation of flat FLRW cosmology, proper time $\tau$ and relational time $\phi$ are related by $\dv{\phi}{\tau} = \pb{\phi}{\mathcal{C}} = \frac{N\pi_{\phi}}{V}$, where $N$ is the lapse and $\mathcal{C}$ is the Hamiltonian constraint. Given the Friedmann equations with respect to proper time $\tau$, their formulation in terms of the relational clock $\phi$ is immediate. For more details, we refer to the appendices of~\cite{Oriti:2016qtz,Marchetti:2020umh}.}
\begin{align}
\left(\frac{V'}{V}\right)^2 & = 12\pi G,\\[7pt]
\frac{V''}{V} & = 12\pi G,
\end{align}
the solution of which is given by
\begin{equation}\label{eq:solution of Friedmann equations}
V(\phi_0)
=
V_0 \exp(\pm\sqrt{12\pi G}\phi_0),
\end{equation}
where $V_0>0$ is determined by initial conditions. 

Following~\cite{Marchetti:2020qsq}, the classical limit as defined by the conditions in Eq.~\eqref{eq:conditions of classical limit} needs to be accompanied by small quantum fluctuations of relevant operators, such as $\hat{N}$, $\hat{V}$ and the scalar field operators. As we demonstrate in Section~\ref{subsec:Phenomenological Implications} in the paragraph on classicalization, the relative volume fluctuations (and implicitly also the relative particle number fluctuations) vanish at late times assuming that there is a phase of dominant-$\rho_0$. Furthermore,~\cite{Marchetti:2020qsq} shows that the fluctuations of the scalar field operator and the scalar field momentum operator are basically controlled by $N^{-1}$ and therefore become small at late times. 

We interpret Eq.~\eqref{eq:emergent G} as the emergence of the gravitational constant from the GFT condensate approach, which relates microscopic variables that enter the definition of the GFT kernel and the clock with the macroscopic parameter $G$~\cite{deCesare:2016axk}. This is a common feature in fluid dynamics, where macroscopic parameters that influence the coarse-grained behavior arise as functions of microscopic ones. 

Note, that the Friedmann equations are not obtained if Eq.~\eqref{eq:emergent G} does not hold. A GFT model and a choice of condensate solutions for which this is the case would then not give a (flat) Friedmann universe dynamics in the continuum limit (at least not within the regime of approximations we have considered). However, in Section \ref{subsec:Phenomenological Implications} we study the emergence of a phase of dominant $\rho_0$ at early and late relational times $\phi\rightarrow\pm\infty$. It turns out that, depending on properties of $\mu_{\rho}$ that are going to be specified, only certain values of $\rho_0$ dominate the early and late behavior. Consequently, the requirement of  Eq.~\eqref{eq:emergent G} becomes a (rather mild) condition on such modes only, i.e. $\mu_{\rho_0}^2 = 3\pi G$.

A different possibility of simplifying the volume equations, Eq.~\eqref{eq:effective CPS Friedmann equation 1} and Eq.~\eqref{eq:effective CPS Friedmann equation 2}, is to consider single-label condensates, which we discuss in Section~\ref{subsec:Single-Label Condensates, Quantum Corrected Friedmann Equations and LQC}.

\subsection{Emergent quantum bounce and comparison to LQC}\label{subsec:Emergent Quantum Bounce}

A sufficient but not necessary condition for a bounce is that at least one $\mathcal{Q}_{\rho}$ is non-zero. Even in the case of $\mathcal{Q}_{\rho} = 0,\;\forall\rho\in\R$, a bounce may occur for at least one $\mathcal{E}_{\rho}$ being strictly negative.

For some $\mathcal{Q}_{\rho}$ being non-zero, Eq.~\eqref{eq:eom for r} reveals that $r_{\rho}$ cannot be zero due to an infinite potential barrier at $r_{\rho}=0$. As a consequence, the expectation value of the volume given in Eq.~\eqref{eq:volume expectation value} is bounded from below, leading to an upper bound of the energy density of the scalar field~\cite{Marchetti:2020umh}
\begin{equation}\label{eq:rho_phi}
\rho_{\phi}
=
\frac{\expval{\hat{\pi}_{\phi}}_{\epsilon;\phi_0,\pi_0}^2}{2 \langle\hat{V}\rangle_{\epsilon;\phi_0,\pi_0}^2},
\end{equation}
which we interpret as the occurrence of a quantum bounce.\footnote{In GR, singularities are defined by divergent curvature invariants. Since curvature operators in GFT have not been studied in detail so far, we characterize a singularity by vanishing volume and divergent energy density. Thus, a singularity resolution corresponds to a lower bound on the volume or an upper bound on the energy density.}  This is the same bounce mechanism as in~\cite{Oriti:2016qtz} and~\cite{Marchetti:2020umh}, where the assumption of $Q_j \neq 0$ for some $\SU$-representation label $j$ arises naturally as it corresponds to having a non-vanishing scalar field which is needed for relational evolution and as matter content of the universe.  However, the integration constant $\mathcal{Q}_{\rho}$ defined by Eq.~\eqref{eq:definition of Q} is related differently to the expectation value of the scalar field momentum and the following paragraph intends to provide a physical interpretation of $\mathcal{Q}_{\rho}$.

Following~\cite{Marchetti:2020umh}, there is a relational Hamiltonian $\hat{H}$ which generates translations in the peaking value $\phi_0$. In the current setting, it is given by
\begin{equation}\label{eq:CPS Hamiltonian}
\hat{H}
=
-i\int\dd{\phi}\int\left[\dd{g}\right]^4\left(\hat{\varphi}^{\dagger}(g_v;\phi)\partial_{\phi}\sigma_{\epsilon}(g_v;\phi)-\hat{\varphi}(g_v;\phi)\partial_{\phi}\bar{\sigma}_{\epsilon}(g_v;\phi)\right),
\end{equation}
entering the relational Schr\"odinger equation
\begin{equation}
-i\dv{}{\phi_0}\ket{\sigma_{\epsilon};\phi_0,\pi_0}
=
\hat{H}\ket{\sigma_{\epsilon};\phi_0,\pi_0}.
\end{equation}
Its expectation value with respect to the CPS evaluates in a saddle-point approximation to~\cite{Marchetti:2020umh}%
\begin{equation}
\langle\hat{H}\rangle_{\sigma_{\epsilon};\phi_0,\pi_0}
=
\pi_0 N(\phi_0).
\end{equation}
Compared with the approximate expectation value of the scalar field momentum operator
\begin{equation}
\langle\hat{\pi}_{\phi}\rangle_{\sigma_{\epsilon};\phi_0,\pi_0}
=
\pi_0\left(\frac{1}{\epsilon\pi_0^2-1}+1\right)N(\phi_0)+\int\dd{\rho}\mathcal{Q}_{\rho},
\end{equation}
the term $\int\dd{\rho}\mathcal{Q}_{\rho}$ controls the difference of momentum operator and relational Hamiltonian~\eqref{eq:CPS Hamiltonian} specified by 
\begin{equation}
\int\dd{\rho}\mathcal{Q}_{\rho}
=
\langle\hat{\pi}_{\phi}\rangle_{\sigma_{\epsilon};\phi_0,\pi_0}-\langle\hat{H}\rangle_{\sigma_{\epsilon};\phi_0,\pi_0},
\end{equation}
in the regime where $\epsilon\pi_0^2\gg 1$. Requiring that relational evolution should be generated by the momentum of the scalar field for all values of $N(\phi_0)$~\cite{Marchetti:2020umh},  we conclude that $\int\dd{\rho}\mathcal{Q}_{\rho}$  needs to vanish. Therefore, in the CPS formalism there is no physical obstruction to set all the charges $\mathcal{Q}_{\rho}$ to zero. Since such a setting leads to meaningful relational evolution as explained above, this configuration may be in fact very reasonable. In this case, for a strictly negative $\mathcal{E}_{\rho}$, the effective potential defined by Eq.~\eqref{eq:eom for r with E} is strictly negative as well and therefore, $r_{\rho}$ cannot reach zero. Hence, although through a different mechanism, a bounce can still occur (see also~\cite{deCesare:2016rsf,Pithis:2016cxg}). 

In conclusion, the bounce does not appear generically but depends on the conserved quantities $\mathcal{Q}_{\rho}$ and $\mathcal{E}_{\rho}$. In particular, the clock parameters $\epsilon$ and $\pi_0$ directly enter the bounce equations and may therefore lead to different bounce behavior. 

Despite this remarkable result, it should be stressed that if the bounce occurs, this represents only a mean value result and quantum fluctuations need to be considered in order to check the stability of the bouncing scenario. For a careful analysis, we refer to~\cite{Marchetti:2020qsq}.

To appreciate the result of an emergent bounce in particular for the extended Lorentzian BC GFT condensate approach, we highlight in the following paragraph the differences between the bounce mechanism as explained here and that of Loop Quantum Cosmology.

In the approach of LQC, where the configuration space of gravity is first restricted to homogeneous and isotropic spacetimes and then loop quantized, one assumes that the underlying spin network is given by a graph with a large number of vertices and edges and where the edges are all labelled by the same spin $j_0$.\footnote{For this reason, the single-label condensate that is introduced in the next section mimics this assumption in the GFT formulation and one finds in fact similar behavior.}

While the singularity resolution for GFT condensates is dynamical, the quantum bounce in LQC is already incorporated at the kinematical level, where the state corresponding to zero volume is decoupled~\cite{Banerjee:2011qu}.\footnote{For other classical or quantum mechanisms, leading to bouncing cosmologies, such as the matter bounce scenario~\cite{Finelli:2001sr,Cai:2014jla,Cai:2015vzv}, the ekpyrotic scenario~\cite{Khoury:2001wf} and string gas cosmology~\cite{Brandenberger1989,Brandenberger:2007zza}, see~\cite{Brandenberger:2016vhg} for a status report.} This is motivated by the cylindrical consistency condition of LQG, which excludes the representation labels $j=0$~\cite{Thiemann2007a}. Since its spectrum is discrete, there is then a finite minimal eigenvalue of the volume operator. No condition such as cylindrical consistency exists a priori in the GFT setting. Furthermore, and most importantly, the spectra of geometric operators in the BC GFT model are continuous. Therefore, we conclude that the appearance of the bounce does not rely on the discreteness of spectra of geometric operators, but is grounded on the non-vanishing number density, which is a consequence of either one non-vanishing term $Q_{\rho}$ or a strictly negative $\mathcal{E}_{\rho}$. More than that, the expectation value of the number operator is required to be not too small, because otherwise, the hydrodynamic approximation ceases to be valid. This in turn can be seen in the behavior of the relative volume fluctuations given in the paragraph on classicalization in Section~\ref{subsec:Phenomenological Implications}, which diverges for vanishing number density.

Apart from the bounce mechanism, LQC and BC condensate cosmology are conceptually very different. First, LQC cannot be seen as the cosmological sector of LQG, since the symmetry reduction was imposed before quantization using loop techniques~\cite{Wilson-Ewing:2016yan}. GFT condensate cosmology on the other hand is in fact a part of the full framework of GFTs since one considers the adaptation to cosmology on the quantum level, as discussed in Sections \ref{subsec:Coherent Peaked Condensate States} and \ref{subsec:Isotropic Restriction and Fixing of Left Intertwiner}. Second, the spin foam models of BC and EPRL are quantizations of differently constrained $BF$-actions. In particular, the EPRL model relies on the modification of the first order Palatini formulation of GR by the Holst term introducing a new parameter, the Barbero-Immirzi parameter $\gamma$.\footnote{Note that in the first order formulation tetrad and connection are independent fields. In the presence of fermions this leads to torsion already at the classical level~\cite{Freidel:2005sn,Perez:2005pm} which, however, is observationally irrelevant at subplanckian energies.} Furthermore, before loop quantization, a $3+1$ split of spacetime with foliation normal vector field $X$ is introduced, reducing the full Lorentz group to the subgroup $\SU_X\subset\SL$. As a result, differences regarding the spectra of geometric operators, the area gap and the representation theory arise, which have been discussed above.

From the above perspective, and despite the underlying spin foam framework and the (Lorentz) spin network states, the BC condensate cosmology approach could be seen as more closely related to the Wheeler-de Witt (WdW) quantization of homogeneous and isotropic spacetimes than to LQC, since the extended Lorentzian BC GFT model is the covariant quantization of the Palatini action and not of the Palatini-Holst action. Thus, the actions on which WdW and BC GFT quantization are based are the same, written in different variables. However, the straightforward quantization of a flat FLRW configuration using WdW techniques leads to no generic singularity resolution, as discussed in~\cite{Kiefer2012,Hamber2009}. It would be interesting to compare this result with the non-genericity of the bounce we obtained in the GFT cosmology setting.

In the next section, we introduce a simplification of the condensate equations of motion that allows for a direct comparison between the quantum bounce arising in GFT condensate cosmology and Loop Quantum Cosmology.

\subsection{Single-Label Condensates}\label{subsec:Single-Label Condensates, Quantum Corrected Friedmann Equations and LQC}

In this section, we consider reduced condensate wavefunctions that are peaked on a single $\SL$-representation label $\rho_0$, i.e. $\sigma_{\rho} = 0,\quad\forall \rho\neq \rho_0$ which is motivated by the emergence of dominant-$\rho_0$ phases, studied below. Following~\cite{Oriti:2016qtz}, not only quantum corrected Friedmann equations are obtained in this approximation, but also the connection of the GFT condensate cosmology approach and LQC~\cite{Bojowald2008,Ashtekar:2021kfp,Banerjee:2011qu} is apparent.

For the correct early and late time behavior, when the condensate is supposed to settle into an FLRW-like state, we impose $\mu_{\rho_0} = 3\pi G$. Under this condition, Eqs.~\eqref{eq:effective CPS Friedmann equation 1} and~\eqref{eq:effective CPS Friedmann equation 2} simplify significantly~\cite{Marchetti:2020umh}
\begin{align}
\left(\frac{V'}{3V}\right)^2
& =
\frac{4\pi G}{3}-\frac{4V_{\rho_0}^2\mathcal{Q}_{\rho_0}^2}{9V^2}+\frac{4V_{\rho_0}\mathcal{E}_{\rho_0}}{9V},\label{eq:single spin Friedmann equation 1}\\[7pt]
\frac{V''}{V}
& =
12\pi G +\frac{2V_{\rho_0}\mathcal{E}_{\rho_0}}{V}.\label{eq:single spin Friedmann equation 2}
\end{align}
For stronger resemblance with the Friedmann equations of LQC shown below in Eqs.~\eqref{eq:LQC Friedmann 1} and~\eqref{eq:LQC Friedmann 2}, using the definition of the energy density of the scalar field in Eq.~\eqref{eq:rho_phi}, we re-write Eq.~\eqref{eq:single spin Friedmann equation 1} as
\begin{equation}\label{eq:single spin Friedmann with rho}
\left(\frac{V'}{3V}\right)^2
 =
\frac{4\pi G}{3}\left(1-\frac{\rho_{\phi}}{\rho_c}\right)
-
\frac{\pi_0^2}{2V_{\rho_0}^2}+\frac{1}{V}\left(\frac{4V_{\rho_0}\mathcal{E}_{\rho_0}}{9}-\pi_0\mathcal{Q}_{\rho_0}\right),
\end{equation}
where $\rho_{\phi}$ was defined in Eq.~\eqref{eq:rho_phi} and we introduced $\rho_c \defeq \frac{3\pi G}{2V_{\rho_0}^2}$, being the critical energy density. Similar to above, the requirement that relational evolution is governed on average by the scalar field momentum results in the condition $\mathcal{Q}_{\rho_0} = 0$, simplifying Eq.~\eqref{eq:single spin Friedmann equation 1} to
\begin{equation}\label{eq:simplified single spin Friedmann equation}
\left(\frac{V'}{3V}\right)^2
=
\frac{4\pi G}{3}+\frac{4V_{\rho_0}\mathcal{E}_{\rho_0}}{9V}.
\end{equation}
Clearly, also in the case of a single-label condensate a bounce can still occur if $\mathcal{Q}_{\rho_0} = 0$, provided that $\mathcal{E}_{\rho_0}$ is strictly negative. Note that due to the additional terms in Eq.~\eqref{eq:single spin Friedmann with rho} containing $\mathcal{E}_{\rho_0}$, $V_{\rho_0}$ and $\pi_0$, the bounce does not occur at $\rho_c$ but is either shifted to larger or lower energy densities depending on the relative magnitudes of the parameters. The term $V_{\rho_0}\mathcal{E}_{\rho_0}$ scales as $\sim\sqrt{\hbar}$ and thus both Eqs.~\eqref{eq:simplified single spin Friedmann equation} and~\eqref{eq:single spin Friedmann equation 2} are the classical Friedmann equations corrected by terms arising from GFT quantum gravity effects~\cite{Oriti:2016qtz}. Remarkably, the quantum corrected equations bear resemblance with the equations obtained in LQC
\begin{align}
\left(\frac{V'}{3V}\right)^2    &= \frac{4\pi G}{3}\left(1-\frac{\rho_{\phi}}{\rho_c^{\text{LQC}}}\right),\label{eq:LQC Friedmann 1}\\[7pt]
\frac{V''}{V}                   &= 12\pi G,\label{eq:LQC Friedmann 2}
\end{align}
where $\rho_c^{\text{LQC}} \sim \rho_{\text{Pl}}/\gamma^3$ is proportional to the Planck energy density~\cite{Ashtekar:2006wn}. Clearly, the quantum corrected equations arising from the GFT condensate approach are similar to those of LQC modulo terms of the order $\sim\sqrt{\hbar}$. As explained above in Section~\ref{subsec:Emergent Quantum Bounce}, despite its similar appearance, the mechanism behind the bounce is very different in LQC and GFT     cosmology.

\subsection{Phenomenological implications}\label{subsec:Phenomenological Implications}

In this section we list the phenomenological implications of the BC GFT condensate cosmology and phenomenological variants of it. 

\paragraph{Dominant-$\rho_0$ phase} 

For EPRL-like condensate cosmology models, the emergence of a phase of dominant representation label has been studied in~\cite{deCesare:2017ynn,Gielen:2016uft,Pithis:2016cxg}. Most importantly, this forms the mechanism behind classicalization and dynamical isotropization~\cite{Pithis:2016cxg}, both of which are discussed below. 

We study the emergence of a phase of a dominant representation label, referred to as dominant-$\rho_0$ phase for brevity, by solving directly the effective equations of motion for the reduced condensate wavefunction~\eqref{eq:approximate eom}. Following a simple exponential ansatz, the solutions of this equation are given by
\begin{equation}
\sigma_{\rho}(\phi_0)
=
\e^{i\tilde{\pi}_0}\left(\alpha_{\rho}^+ \e^{\mu_{\rho}\phi_0} + \alpha_{\rho}^- \e^{-\mu_{\rho}\phi_0}\right),
\end{equation}
where $\alpha_{\rho}^{\pm}\in\C$ are constants of integration and the phase factor $\e^{i\tilde{\pi}_0}$ leads to a factor of unity for any physical observable, since these depend only on the modulus of $\sigma_{\rho}$.

Clearly, the values of $\mu_{\rho}$ for different labels $\rho$ determine the behavior of the reduced condensate wavefunction. Assuming that $\mu_{\rho}$ has a positive global maximum at $\rho_0$\footnote{If the kinetic kernel consists of a Laplacian and the coupling constants have appropriate signs, this assumption is satisfied. For instance, this is the case for the example studied in~\cite{Gielen:2016uft}.}, the reduced condensate wavefunction is quickly dominated by $\sigma_{\rho_0}$. Therefore, the expectation value of the volume operator takes the following form in the limit of early and late times
\begin{equation}
\langle\hat{V}\rangle_{\epsilon;\phi_0,\pi_0}
\underset{\phi_0\rightarrow\pm\infty}{\rightarrow}V_{\rho_0}\abs{\alpha_{\rho_0}^{\pm}}^2\e^{\pm 2\mu_{\rho_0}}
\end{equation}
For $\mu_{\rho}$ given explicitly by
\begin{equation}
\mu_{\rho}
=
\frac{\pi_0^2}{\epsilon\pi_0^2-1}\left(\frac{2}{\epsilon\pi_0^2}-\frac{1}{\epsilon\pi_0^2-1}\right)+\frac{B_{\rho}}{A_{\rho}},
\end{equation}
we follow that the global positive maxima of $\mu_{\rho}$ are determined by those of $B_{\rho}/A_{\rho}$. Hence, the early and late time behavior of the condensate is determined by the explicit form of the kinetic kernels that enter the GFT action.

Consistency with the macroscopic Friedmann equations requires then, that asymptotically,
\begin{equation}
\mu_{\rho_0} = 3\pi G
\end{equation} 
Consequently, one finds an effective gravitational constant $G_{\text{eff}}(\phi_0)$ that attains the value of the Newtonian gravitational $G$ only in the limit $\phi_0\rightarrow\pm\infty$, analogous to the study given in~\cite{deCesare:2016rsf}. For all other values of $\rho\neq\rho_0$, no condition is imposed on $\mu_{\rho}$ in contrast to the too restrictive case considered in Eq.~\eqref{eq:emergent G}. 

\paragraph{Classicalization}

Given that a dominant-$\rho_0$ phase emerges, we adapt the discussion on classicalization in~\cite{Pithis:2016cxg} to the reduced condensate wavefunction labelled by the continuous $\SL$-parameter $\rho$. Classicalization at early and late times means that volume fluctuations compared to the expectation value of the volume go to zero for $\phi_0\rightarrow\pm\infty$. In analogy to~\cite{Pithis:2016cxg}, we define
\begin{equation}
\frac{\delta V}{V}
\defeq
\frac{\sqrt{\langle\hat{V}^2\rangle_{\epsilon;\phi_0,\pi_0}-\langle\hat{V}\rangle_{\epsilon;\phi_0,\pi_0}^2}}{\langle\hat{V}\rangle_{\epsilon;\phi_0,\pi_0}}.
\end{equation}
Using the commutation relation in Eq.~\eqref{eq:commutation relations of group field}, one can easily show that
\begin{equation}
\langle\hat{V}^2\rangle_{\epsilon;\phi_0,\pi_0}
=
\langle\hat{V}\rangle_{\epsilon;\phi_0,\pi_0}^2+\int\dd{\rho}V_{\rho}^2\abs{\sigma_{\rho}(\phi_0)}^2.
\end{equation}
Hence, the relative fluctuations vanish in the limit of early and late times
\begin{equation}
\frac{\delta V}{V}
\underset{\phi\rightarrow\pm\infty}{\longrightarrow}\frac{\left(V_{\rho_0}^2\abs{\alpha_{\rho_0}^{\pm}}^2\e^{\pm 2 \mu_{\rho_0}}\right)^{1/2}}{V_{\rho_0}\abs{\alpha_{\rho_0}^{\pm}}^2\e^{\pm 2 \mu_{\rho_0}}}\longrightarrow 0.
\end{equation}

\paragraph{Dynamical isotropization}

Another effect, the mechanism of which is based on the emergence of a dominant-$\rho_0$ phase is dynamical isotropization. We briefly present it in the following, based on the discussion in~\cite{Pithis:2016cxg} for $\SU$-data. Suppose that the assumption of isotropy is dropped and we therefore consider reduced condensate wavefunctions $\sigma_{\rho_1 \rho_2 \rho_3 \rho_4}(\phi_0)$ depending on four possibly different values of $\rho_i$. Then, in the approximation of small simplicial interactions, the early and late time behavior of the condensate is dictated by the kinetic kernels $\mathscr{K}_{(i)}^{\rho_1 \rho_2 \rho_3 \rho_4}$ which enter the definition of an anisotropic $\mu_{\rho_1 \rho_2 \rho_3 \rho_4}$, defined similarly to $\mu_{\rho}$. If $\mu_{\rho_1 \rho_2 \rho_3 \rho_4}$ exhibits a positive global maximum at $(\rho_0,\rho_0,\rho_0,\rho_0)$ where all the $\rho_i$ are equal, then $\sigma_{\rho_i}(\phi_0)$ will be quickly dominated by $ \sigma_{\rho_0\rho_0 \rho_0 \rho_0}$ for $\phi_0\rightarrow\pm\infty$. For instance, this condition is met in case of the kinetic kernel consisting of a Laplacian, similar to the example in~\cite{Pithis:2016cxg}. 

Bringing the three preceding paragraphs together, the following picture emerges: based on the mechanism of emerging low-spin phases, an anisotropic reduced condensate wavefunction will rather generally isotropize dynamically at late times, resulting in a flat FLRW spacetime for $\mu_{\rho_0 \rho_0 \rho_0 \rho_0} = 3\pi G$ with vanishing relative volume fluctuations.

\paragraph{Quantum geometric inflation}

Besides big bang resolution, emergent Friedmann equations and dynamical isotropization, phenomenological aspects of GFT condensate cosmologies leading to an accelerated expansion have been considered for EPRL-like single spin condensates~\cite{deCesare:2016rsf,Pithis:2016cxg}. The idea behind the study of what can be phrased \enquote{quantum geometric inflation} is not only that the inflaton is replaced by quantum gravity effects but also that strong constraints can be put on the microscopic GFT parameters in order to match current observations and approved aspects of standard cosmology. 

Consider a single-label BC condensate $\sigma(\phi_0)=r(\phi_0)\e^{i\theta(\phi_0)}$ peaked on a label $\rho_0$ with volume $V = V_{\rho_0}r^2$, conserved charge $\mathcal{Q}$ and conserved $\mathcal{E}$, governed by an effective action the interaction term of which models pseudo-simplicial and pseudo-tensorial interactions.\footnote{Considering the potential of the Barrett-Crane GFT formulation defined in \eqref{eq:divergent BO action}, we find that the simplicial interactions terms are not bounded from below and break the global $\text{U}(1)$-symmetry.  Therefore, we emphasize that the potential only mimicks simplicial and tensor invariant interactions, which are thus referred to as pseudo-simplicial and pseudo-tensorial interactions~\cite{deCesare:2016rsf}.} Arising from the effective action, the equation of motion for the radial part of the reduced condensate wavefuntion is given by
\begin{equation}\label{eq:phenomenological eom}
r''-\mu^2 r-\frac{\mathcal{Q}^2}{r^3}+\alpha r^{n-1}+\beta r^{n'-1} = 0,
\end{equation}
where
\begin{equation}
\beta > 0,\quad \mu^2 >0,\quad \alpha,\quad n'>n
\end{equation}
are microscopic parameters entering the effective action. Notice that they not only depend on kernels of the GFT action but also on the clock parameters $\epsilon$ and $\pi_0$. Eq.~\eqref{eq:phenomenological eom} is equal to that of a point particle in an effective potential that admits cyclic universes, studied in detail in~\cite{deCesare:2016rsf}. 

Acceleration of the universe $\mathfrak{a}$\footnote{Since $\ddot{a}\propto V''/V-\frac{5}{3}(V'/V)^2$, the sign of $\ddot{a}$ is captured by $\mathfrak{a}\defeq V''/V-\frac{5}{3}(V'/V)^2 $.} , i.e. positive second derivative of the scale factor, can be expressed purely in relational terms by
\begin{equation}\label{eq:phenomenological acceleration}
\mathfrak{a}(r)
=
-\frac{2}{3r^4}\left[P(r)+\gamma r^{n+2}+\delta r^{n'+2}\right],
\end{equation}
where we introduced
\begin{equation}
P(r) \defeq 4\mu^2r^4+14\mathcal{E}r^2-10\mathcal{Q}^2, \quad \gamma \defeq \left(3-\frac{14}{n}\right)\alpha,\quad\delta \defeq \left(3-\frac{14}{n'}\right)\beta.
\end{equation}
Studying first the free case where interactions are absent, i.e. $\gamma = \delta = 0$, the number of e-folds $N$ between bounce and end of inflation
\begin{equation}\label{eq:number of efolds}
N
=
\frac{1}{3}\log(\frac{V_{\text{end}}}{V_{\text{bounce}}})
=
\frac{2}{3}\log(\frac{r_{\text{end}}}{r_{\text{bounce}}}),
\end{equation}
is bounded
\begin{equation}
\frac{1}{3}\log(\frac{10}{3})\leq N \leq 
\frac{1}{3}\log(\frac{7}{4}).
\end{equation}
Following standard cosmological literature on inflation~\cite{Mukhanov2005}, the number of e-folds $N$ must satisfy $N\gtrsim 60$ for the stage between the bounce and the end of inflation to be called an inflationary era. Hence, we conclude that the free condensate does not admit a phase of accelerated expansion and therefore no quantum geometry-induced inflation. 

Consequently, incorporating interactions is necessary. The line of arguments presented in~\cite{deCesare:2016rsf} is the following. In order to have an extended phase of cosmological expansion, the near-bounce behavior should be dominated by the free theory and sub-leading terms while the condensate at late times, where many GFT quanta are present, should be dominated solely by the contribution $\sim r^{n'}$. Therefore, we impose
\begin{equation}
\beta\ll\abs{\alpha},
\end{equation}
which leads to an extension of the time between bounce and end of inflation. Requiring in addition that the universe should have Planckian volume at the bounce, the parameters $\alpha,\mathcal{Q}^2$ and $\mu^2$ are subject to the following inequality~\cite{deCesare:2016rsf}
\begin{equation}
\alpha\leq \mu^2+\mathcal{Q}^2.
\end{equation}
For an extended period of acceleration, we need to demand $\gamma<0$ as otherwise, the end of inflation is reached even earlier than in the case of the free theory~\cite{deCesare:2016rsf}. This implies that $n'>\frac{14}{3}$. The first integer $n'$ to match this inequality is given by $5$ which mimicks simplicial interactions in four dimension. Furthermore, we set $\alpha<0$ so that there's no intermediate phase of deceleration. Higher order terms with power $n'>5$ could account for interactions generated in the RG flow. If $n'$ is even, the highest order term could represent tensor-invariant interactions~\cite{deCesare:2016rsf}.

In conclusion, these arguments demonstrate that phenomenological considerations can be used to restrict the possible GFT coupling and clock parameters and pose conditions among them.

\paragraph{Phantom-like dark energy}

As a last phenomenological aspect being also the most recent one, we transfer the results of~\cite{Oriti:2021rvm} to the extended Lorentzian BC GFT condensate cosmology. Based on a similar phenomenological model than the one presented in the preceding paragraph,~\cite{Oriti:2021rvm} studies the effects of a condensate with two contributing modes, rather than just single-label condensates. Moreover, instead of insisting on an inflation like phase in the early universe, the focus is on the possibility to have a late-time accelerated phase of expansion, following the bounce (which could in principle provide in itself a viable scenario for the early universe inhomogeneities and then structure formation) and a long-lasting Friedmann evolution. In other words, the role of GFT interactions and quantum gravity in general would be to explain dark energy as a macroscopic aspect of an emergent spacetime scenario. 

First, before moving to the late-time behaviour, the behavior of the condensate close to the bounce is studied, where the free part is expected to be dominant while the interactions are negligible due to the smallness of $\sigma_{\rho}$. Similar to the single-label condensate studies~\cite{deCesare:2016rsf}, an upper bound on the number of e-folds is found in the more general setting of a two mode condensate, therefore not exhibiting a long-lasting phase of inflation in the free case~\cite{Oriti:2021rvm}. Then, the condensate either settles into an FLRW phase or leads to a cyclic universe, depending on the point where interactions become non-negligible.

If the FLRW phase has been reached,~\cite{Oriti:2021rvm} shows that the behavior of the condensate drastically depends on whether one or two modes contribute.

In the case of a single-label condensate, the equation of state, determined in relational terms via
\begin{equation}
w = 3-\frac{V V''}{V'^2},
\end{equation}
is asymptotically determined by the degree $n$ of interaction in the phenomenological interaction term. In particular, the universe ends in a big rip singularity for $n>6$~\cite{Oriti:2021rvm}.

In contrast, the late time behavior for a two-mode condensate is much more intricate. Remarkably, the barrier $w=-1$ is crossed to lower values, leading to a time of phantom dark energy with equation of state $w<-1$. Asymptotically, $w=-1$ is reached from below without hitting a future singularity. Instead, one finds a de Sitter-like spacetime at late times~\cite{Oriti:2021rvm}. 

\section{Discussion and Conclusion}\label{sec:Discussion and Conclusion}

The goal of this article was to explore kinematical and dynamical aspects of condensate cosmology in the extended Lorentzian BC GFT model based on a timelike normal and spacelike tetrahedra. As a first result we obtained the detailed spin representation of this model, generalizing the results for the Riemannian model in~\cite{Baratin:2011tx}. This provided the basis for the subsequent analysis. On the kinematical side, we clarified the relation between the minisuperspace of homogeneous spatial cosmologies and the domain of the GFT condensate wavefunction in this model. We showed that they are indeed diffeomorphic, serving as an important consistency check of the cosmological interpretation (and in fact a main motivation) for GFT condensate cosmology as a whole~\cite{Gielen:2014ila,deCesare:2017ynn}. This is our second main result.  
Then we derived the emergent cosmological evolution of the extended Lorentzian BC GFT model, specifically the effective dynamics of the universe volume, and demonstrated that it is very similar to that of the EPRL-like GFT model. This is our third main result. 
In fact, the resulting equations basically coincide upon identification of the $\SU$-representation index $j$ with the $\SL$ index $\rho$. If at the formal level the resulting matching of cosmological dynamics, including the quantum bounce, is then not surprising, this shows that the underlying quantum gravity mechanism producing such quantum bounce is not tied to the discreteness of geometric spectra nor to the value of the Barbero-Immirzi parameter (which does not appear at all in our model). This is to be contrasted with the situation in loop quantum cosmology, for example, despite the similarity of the respective cosmological equations, and with the general expectation in canonical loop quantum gravity. Clarifying this point is another interesting aspect of our work.
\newline
\newline\indent In order to appreciate our findings at the cosmological level, i.e. both the similarity in the effective dynamics and the appearance of the cosmic bounce, it is worthwhile to highlight the main differences between the EPRL-like and BC GFT models. First, the Barbero-Immirzi parameter $\gamma$ enters the simplicity constraint and the spectra of operators in the EPRL-like models, and is interpreted as an irreducible quantization ambiguity in canonical LQG. Its presence implies a propensity to break parity~\cite{Ashtekar:1988sw,Freidel:2005sn,Contaldi:2008yz} and raises questions regarding its value and its running as a coupling constant~\cite{Benedetti:2011yb,Charles:2016mjn}. These points do not concern the Barrett-Crane model, which is the spin foam or GFT quantization of the Palatini action. In particular, the spectra of geometric operators in the BC model do not depend on $\gamma$. Moreover, the area spectrum for triangles (and, we expect, for other spatial geometric operators) in the BC model is continuous, but with a natural area gap (in absence of any requirement of cylindrical equivalence on the corresponding states). Second, the BC GFT model explicitly uses the the full Lorentz group $\SL$ and does not allow for the usual map to a formulation in terms of boundary states and GFT fields using only the rotation subgroup $\SU$~\cite{Finocchiaro:2020xwr}, heavily used in the EPRL context. Last, the extended BC GFT model is unambiguous in its definition, since the composite geometricity operator imposing both simplicity and closure conditions is a projector; thus one cannot expect that variations of the fundamental construction would lead to results different from ours, and substantially different from the EPRL ones.  

In the light of these differences, how do we interpret the equivalent effective cosmological evolution of the two types of GFT models? It seems to us that the most reasonable perspective is that of universality of the continuum limit. In other words, our results serve as a reminder that the coarse-graining necessarily involved in extracting continuum physics from fundamentally discrete quantum gravity models could be quite powerful. That is, they lend support to the conjecture that the EPRL-like and BC GFT model lie in the same universality class, from the point of view of continuum gravitational physics. 

We note that this is consonant to the recent work at the lattice level~\cite{Dittrich:2021kzs}, showing that, at least for hypercubic lattices, the microscopic details of spin foam models such as the precise implementation of simplicity constraints do not affect the behavior on large scales.
\newline
\newline\indent Clearly, to establish our results, several approximations were needed and thus such results can only be trusted within their domain of validity. This limits also the solidity of our interpretation in terms of universality class. The coincidence of effective cosmological dynamics between BC and EPRL models may well be broken as soon as our approximations are relaxed, in principle.

Besides, when the GFT \enquote{particle number} grows too much, we cannot expect the coherent condensate states to be a quantitatively good approximation of the true ground state of the quantum gravity system (they do not approximate well exact solutions of the quantum equations of motion). The same is true, more generally, when GFT interactions are not subdominant to the free theory. This is apparent in the fact that simple condensate states are product states, which do not encode any connectivity information among the simplices dual to GFT quanta, i.e. they do not encode any entanglement among them. Moreover, the restriction to isotropy imposed directly by symmetry reduction at the level of the condensate wavefunction may be too strong (note for example that this leads to a complete decoupling of different field modes in the mean-field dynamics, in both models). One could consider imposing it as well in terms of coarse-graining of anisotropic data. Finally, we restricted our attention to condensate wavefunctions with fixed intertwiner labels, as we argued that this is sufficient, if not needed, to capture the correct intrinsic geometry of the BC tetrahedra (and thus of the effective continuum spatial geometry) in the isotropic case. Our argument should be further supported by some explicit evaluation of geometric operators, testing the (ir)relevance of the left $\SL$-intertwiner label. This is also missing, incidentally, in the derivations based on the EPRL-like model~\cite{deCesare:2017ynn}.

In conclusion, we take our present results and those of~\cite{Dittrich:2021kzs} and~\cite{Baratin:2011tx}, to encourage much further work on the continuum regime (as well as on the discrete aspects) of the BC model, and to suggest that its dismissal as a candidate for a fundamental formulation of quantum gravity in four dimensions is premature.

Finally, we would like to emphasize that the Feynman diagrams of both the EPRL-like and the BC GFT model considered so far in the context of GFT condensate cosmology are dual to triangulations formed by spacelike tetrahedra only. While this is not necessarily a restriction of the overall spacetime geometry, they still represent a very special class of triangulations of Lorentzian manifolds. Work on extensions of the model including also timelike triangles is given in~\cite{Conrady:2010vx,Conrady:2010kc} for the EPRL spin foam model and in~\cite{Barrett:1999qw,Perez:2000ep} for the Barrett-Crane model. A priori the inclusion of timelike (or null) tetrahedra may not be needed to encode cosmological evolution, since it is not required to specify quantum states of the universe or relevant relational observables. However, the causal structure is certainly relevant for correct (unitary) encoding of the (quantum) spacetime dynamics. Thus, we suspect that a (potentially colored~\cite{Gurau:2011xp}) GFT model, which explicitly incorporates all the possible gluings of spacelike, timelike and null faces is needed to encode more completely the discrete and quantum seeds of continuum causal structures (in a similar way as done in CDT~\cite{Ambjorn:2001cv,Gorlich:2013hu,Ambjorn:2013tki,Loll:2019rdj}) and might thus exhibit a different phase structure. Furthermore, only in such a complete model it is possible to define timelike and null boundary states, which is important to give a quantum geometric description of anti-de Sitter space, cosmological and black hole horizons (see also~\cite{Speziale:2013ifa}), as well as physical propagation of inhomogeneities. This will then be an important next research direction, to develop further our work. 

\subsection*{Acknowledgments}

The authors thank Luca Marchetti for helpful comments on the manuscript. D. Oriti and A. Pithis acknowledge funding from DFG research grants OR432/3-1 and OR432/4-1.
The work of A. Pithis leading to this publication was also supported by the PRIME programme of the German Academic Exchange Service (DAAD) with funds from the German Federal Ministry of Education and Research (BMBF). The authors thank an anonymous referee for comments which improved the manuscript.

\appendix

\section{Representation Theory of $\SL$}\label{appendix:Representation Theory of SL2C}

In the following, we give a synopsis of $\SL$-representation theory based on~\cite{Duc1967,Martin-Dussaud:2019ypf,Ruehl1970} which tailored to the needs of this article.

The special linear group $\SL$ is defined as the set
\begin{equation}
\SL\defeq\left\{g = \mqty(g_{11} & g_{12}\\ g_{21} & g_{22})\; \Bigg{\vert} \; g_{ij}\in\C,\; \det(g) = g_{11}g_{22}-g_{12}g_{21} = 1\right\},
\end{equation} 
with group multiplication being matrix multiplication. The differential structure inherited from $\C^4$ turns it into a non-compact six-dimensional Lie group. Equipped with the Haar measure, denoted by $\dd{g}$, yields the space of square-integrable functions on $\SL$, referred to as $L^2(\SL)$. 

Among the many subgroups of $\SL$, there are two unitary three-dimensional subgroups, given by $\SU$ and SU$(1,1)$. Forming the quotient space with respect to these groups, i.e. $\SL/\SU$ and $\SL/$SU$(1,1)$ yields homogeneous spaces on which $\SL$ acts transitively. When imposing the simplicity constraint as in Eq.\eqref{eq:generalized simplicity}, the homogeneous space $\SL/\SU$ is of particular interest. Since $\SL/\SU$ and the 3-hyperboloid $\HH$ are diffeomorphic, we write equivalently $X\equiv [a]\in\HH$ for $a\in\SL$ a representative. The transitive left action of $\SL$ on $\HH$ is explicitly defined by
\begin{equation}
h\cdot X \equiv h\cdot[a] \defeq [ha],
\end{equation}
from which the definition of the stabilizer subgroup $\SU_X\subset\SL$ follows immediately
\begin{equation}\label{eq:invariance under seperate SU2}
\SU_X
\defeq
\left\{ h\in\SL \; \vert \; h\cdot X = X\right\}.
\end{equation}

The Lie algebra of $\SL$, denoted by $\spl$, is the vector space of traceless $2\times 2$ complex matrices equipped with the Lie bracket being the commutator. The generators of rotations $L_a \defeq \frac{1}{2}\sigma_a$ and boosts $K_a \defeq \frac{i}{2}\sigma_a$, $\sigma_a$ being the Pauli matrices, form a basis of $\spl$ and satisfy the commutation relations
\begin{equation}\label{eq:commutation relations of SL-generators}
\comm{L_a}{L_b} = i\sum_c \varepsilon_{abc}L_c,\quad \comm{L_a}{K_b} = i\sum_c\varepsilon_{abc}K_c,\quad \comm{K_a}{K_b} = -i\sum_c\varepsilon_{abc}L_c.
\end{equation}

Since $\SL$ is simply connected and non-compact, the finite irreducible representations are not unitary. Thus, we turn our attention to the infinite irreducible representations of $\SL$~\cite{Martin-Dussaud:2019ypf}. These can be realized on the space of homogeneous functions of degree $(\lambda,\mu)\in\C^2$ on $\C^2$, denoted by $\mathcal{D}^{(\lambda,\mu)}[z_0,z_1]$. The action of $\SL$ on this space is defined by
\begin{equation}
\begin{aligned}
\mathbf{D}^{(\lambda,\mu)}: \SL & \longrightarrow \text{End}\left(\mathcal{D}^{(\lambda,\mu)}[z_0,z_1]\right)\\[7pt]
g & \longmapsto \left(\mathbf{D}^{(\lambda,\mu)}(g)F\right)(\vb{z})\defeq F(g^{\text{T}}\vb{z})
\end{aligned}
\end{equation}
for $F\in\mathcal{D}^{(\lambda,\mu)}[z_0,z_1]$. We require the irreducible representations to be unitary which yields two distinct conditions on the degrees $(\lambda,\mu)$, the principal series and the complementary series. We are going to focus on the principal series, characterized by $\lambda = i\rho+\nu-1$ and $\mu =i\rho-\nu-1$, where $\rho\in\R$ and $\nu\in\frac{\mathbb{Z}}{2}$. Let $\mathcal{D}^{(\rho,\nu)}[z_0,z_1]$ denote the representation spaces of the principal series.

The map 
\begin{align*}
\kappa: \SU & \longrightarrow \C^2\\[7pt]
u & \longmapsto (u_{21}, u_{22}),
\end{align*}
induces an isomorphic intertwiner $\kappa^{*}$ between homogeneous functions on $\C^2$ and homogeneous functions on $\SU$. One can show that it satisfies the so-called covariance property~\cite{Martin-Dussaud:2019ypf}, i.e. for $F\in\mathcal{D}^{(\rho,\nu)}[z_0,z_1]$ and $\theta\in\R$,
\begin{equation}\label{eq:covariance condition}
\kappa^{*}F\left(\e^{i\theta\sigma_3}u\right) = \e^{-2i\theta \nu}\kappa^{*}F(u).
\end{equation}
It is well-known that the $\SU$-Wigner matrices $D^j_{mn}(u)$ form an orthogonal basis of $L^2(\SU)$. Using the covariance condition \eqref{eq:covariance condition}, we determine the condition for the $\SU$-Wigner matrices such that they live in the space $\mathcal{D}^{(\rho,\nu)}[u]$
\begin{equation}\label{eq:covariance condition on Wigner matrices}
D^j_{mn}\left(\e^{i\theta\sigma_3}u\right) = \e^{2i\theta m}D^j_{mn}(u) \overset{!}{=} \e^{-2i\theta \nu}D^j_{mn}(u)\quad\Rightarrow m = -\nu.
\end{equation}
Therefore, the spin $j$ must be at least $\abs{\nu}$, i.e. $j\in\left\{\abs{\nu},\abs{\nu}+1,...\right\}$ and $n\in\left\{-j,...,j\right\}$. As a consequence, the space $\mathcal{D}^{(\rho,\nu)}[u]$ can be decomposed into a direct sum of $\SU$-representations spaces $\mathcal{Q}_j$
\begin{equation}\label{eq:decomposition of SL2C reps}
\mathcal{D}^{(\rho,\nu)}[u] \cong \bigoplus_{j = \abs{\nu}}^{\infty} \mathcal{Q}_j
\end{equation}
Since the $\SU$-Wigner matrices form an orthogonal basis of $L^2(\SU)$, it is easily verified that the canonical basis, defined by the vectors
\begin{equation}
\phi^j_n(u) \defeq (2j+1)D^j_{-\nu n}(u),
\end{equation}
forms an orthogonal basis of $\mathcal{D}^{(\rho,\nu)}[u]$. Note that equivalently, the canonical basis is denoted in the Dirac bra-ket notation as $\ket{(\rho,\nu);jm}$, defined via the relation $\phi^j_m(u) = \braket{u}{(\rho,\nu);jm}$. In accordance with~\cite{Perez:2000ec}, we define the $\SL$-Wigner matrices as
\begin{equation}\label{eq:Definition of SL2C Wigner matrices}
D^{(\rho,\nu)}_{jmln}(g)
\defeq
\int\limits_{\SU}\dd{u}\bar{\phi}^j_m(u) \mathbf{D}^{(\rho,\nu)}(g)\phi^l_n(u),
\end{equation}
which satisfy the orthogonality relation~\cite{Martin-Dussaud:2019ypf}
\begin{equation}\label{eq:orthogonality relation of SL2C wigner matrices}
\int\limits_{\SL}\dd{h}\overline{D^{(\rho_1,\nu_1)}_{j_1 m_1 l_1 n_1}(h)}D^{(\rho_2,\nu_2)}_{j_2 m_2 l_2 n_2}(h)
=
\frac{\delta(\rho_1-\rho_2)\delta_{\nu_1, \nu_2}\delta_{j_1, j_2}\delta_{l_1, l_2}\delta_{m_1, m_2}\delta_{n_1, n_2}}{4\left(\rho_1^2+\nu_1^2\right)},
\end{equation}
and the complex conjugation property~\cite{Speziale:2016axj}
\begin{equation}
\overline{D^{(\rho,\nu)}_{jmln}(g)}
=
(-1)^{j-l+m-n}D^{(\rho,\nu)}_{j-ml-n}(g).
\end{equation}

The defining property of representation matrices is that they are a group homomorphism. Therefore, the representation of a product of two elements $g_1,g_2\in\SL$ is given as the product of two representation matrices, i.e.
\begin{equation}\label{eq:SL2C-Wigner matrix evaluated on product}
D^{(\rho,\nu)}_{jmln}(g_1g_2) 
=
\sum_{j'=\abs{\nu}}^\infty \sum_{m' = -j'}^{j'}D^{(\rho,\nu)}_{jmj'm'}(g_1)D^{(\rho,\nu)}_{j'm'ln}(g_2).
\end{equation}

Using Eq.~\eqref{eq:Definition of SL2C Wigner matrices}, the evaluation of $\SL$-Wigner matrices on $\SU$ elements leads to
\begin{equation}\label{eq:Evaluating SL-Wigner matrix on SU elements}
D^{(\rho,\nu)}_{jmln}(u) = \delta_{jl}D^j_{mn}(u)
\end{equation}
and the integration of these elements over $\SU$ yields
\begin{equation}
\int\limits_{\SU}\dd{u}D^{(\rho,\nu)}_{jmln}(u)
=
\delta_{jl}\delta_{j,0}\delta_{m,0}\delta_{n,0}\delta_{\nu,0}.
\end{equation}
Therein, we notice the appearance of $\delta_{\nu,0}$. Since $j$ is restricted to zero and the range of $j$ is $j\in\{\abs{\nu},\abs{\nu}+1,...\}$, $k$ must be zero for a possibly non-vanishing contribution. This is an important result which shows that the imposition of simplicity with respect to a timelike normal as in Eq.~\eqref{eq:generalized simplicity} is equal to restricting to representations of the form $(\rho,0)$.

Following Eq.~\eqref{eq:Evaluating SL-Wigner matrix on SU elements} and the Cartan decomposition of $\SL$, i.e.
\begin{equation}\label{eq:Cartan decomposition of g}
\forall g\in \SL\;\exists\; u,v\in\SU,r\in \R_{+}: g = u \e^{\frac{r}{2}\sigma_3}v^{-1}
\end{equation}
the $\SL$-Wigner matrix can be decomposed as
\begin{equation}\label{eq:Cartan decomposition of SL-Wigner matrix}
D^{(\rho,\nu)}_{jmln}(g) 
=
\sum_{q = -\min(j,l)}^{\min(j,l)}D^j_{mq}(u)d^{(\rho,\nu)}_{jlq}(r)D^l_{qn}(v^{-1}),
\end{equation}
where we introduced the \textit{reduced} $\SL$-Wigner matrix
\begin{equation}\label{eq:definition of reduced Wigner matrix}
d^{(\rho,\nu)}_{jlq}(r) \defeq D^{(\rho,\nu)}_{jqlq}\left(\e^{\frac{r}{2}\sigma_3}\right).
\end{equation}
In addition, the Cartan decomposition induces a decomposition of the measure according to
\begin{equation}\label{eq:Cartan decomposition of Haar measure}
\dd{g} = \frac{1}{4\pi}\sinh(r)^2\dd{r}\dd{u}\dd{v} \eqdef \dd{\mu(r)}\dd{u}\dd{v}.
\end{equation}

The two Casimir operators of representations of $\spl$ play a crucial role in the study of the simplicity constraints and volume operators. We construct them from the generators of $\SL$  given in Eq.~\eqref{eq:commutation relations of SL-generators}, yielding
\begin{equation}
\begin{aligned}
\cas_1 & = \vb{K}^2-\vb{L}^2\\[7pt]
\cas_2 & = \vb{K}\cdot\vb{L}
\end{aligned}
\end{equation}
The Casimir operators act on the canonical basis by
\begin{align}
\cas_1\ket{(\rho,\nu);jm} & =  \rho^2-\nu^2+1\ket{(\rho,\nu);jm}\label{eq:definition of cas1}\\[7pt]
\cas_2\ket{(\rho,\nu);jm} & = pk\ket{(\rho,\nu);jm}\label{eq:definition of cas2}.
\end{align}

Together with the expansion of a function $f\in L^2\left(\SL^4\right)$ in terms of representation labels
\begin{equation}\label{eq:General expansion of function on SL2C^4}
\begin{aligned}
f(g_1 ,g_2 ,g_3 ,g_4)
& =
\left[\prod_{i=1}^4 \int\displaylimits_{\R}\dd{\rho_i}\sum_{\nu_i\in\frac{\mathbb{Z}}{2}}4\left(\rho_i^2+\nu_i^2\right)\sum_{j_i = \abs{\nu_i}}^{\infty}\sum_{l_i = \abs{\nu_i}}^{\infty}\sum_{m_i = - j_i}^{j_i}\sum_{n_i = -l_i}^{l_i}\right]\times\\[7pt]
& \times f^{\rho_1 \rho_2 \rho_3 \rho_4 \nu_1 \nu_2 \nu_3 \nu_4}_{j_1 m_1 j_2 m_2 j_3 m_3 j_4 m_4 l_1 n_1 l_2 n_2 l_3 n_3 l_4 n_4}\prod_{i=1}^4 D^{(\rho_i,\nu_i)}_{j_i m_i l_i n_i}(g_i),
\end{aligned}
\end{equation}
the formulae \eqref{eq:SL2C-Wigner matrix evaluated on product}-\eqref{eq:General expansion of function on SL2C^4} form the mathematical basis for the computations involving $\SL$-representation theory in this article. In the next appendix, we introduce the necessary notions of $\SL$-recoupling theory required for Section~\ref{subsec:Spin Representation of Condensate Wavefunction} and~\ref{subsec:Spin Representation of the Action and Regularization}.

\section{Recoupling Theory of $\SL$}\label{appendix:Recoupling Theory of SL2C}

In the following, we construct the necessary notions of $\SL$ recoupling theory, mainly based on~\cite{Speziale:2016axj}, which itself is a refinement of the work done in~\cite{Ruehl1970,Naimark1964,Anderson1970,Anderson1970a,Kerimov1978}.

As for $\SU$-recoupling theory, the basic objects are the Clebsch-Gordan (CG) coefficients, which relate basis vectors of the composite system to the basis vector of the decomposed representation. The tensor product of two representation spaces $\mathcal{D}^{(\rho_1,\nu_1)}$ and $\mathcal{D}^{(\rho_2,\nu_2)}$ is decomposed into $\SL$-representations by the relation
\begin{equation}
\mathcal{D}^{(\rho_1,\nu_1)}\otimes\mathcal{D}^{(\rho_2,\nu_2)}
\cong
\int\dd{\rho}\bigoplus_{\nu\vert \nu_1+\nu_2+\nu\in\mathbb{Z}}\mathcal{D}^{(\rho,\nu)}.
\end{equation} 
In terms of basis vectors $\ket{(\rho,\nu);jm}$, this decomposition is characterized by the $\SL$-CG coefficients $C^{\rho\nu jm}_{\rho_1\nu_1 j_1 m_1, \rho_2\nu_2 j_2 m_2}$ via
\begin{equation}
\ket{(\rho_1,\nu_1);j_1 m_1}\otimes\ket{(\rho_2,\nu_2); j_2 m_2}
=
\int\dd{\rho}\sum_{\nu}\sum_{j = \abs{\nu}}^{\infty}\sum_{m = -j}^{j} C^{\rho\nu jm}_{\rho_1\nu_1 j_1 m_1, \rho_2\nu_2 j_2 m_2} \ket{(\rho,\nu);jm},
\end{equation}
where the $\SU$-spin variables are subject to the usual triangle inequalities
\begin{equation}
\abs{j_1 - j_2}\leq j \leq j_1 + j_2,
\end{equation}
and $\nu$ is constrained by
\begin{equation}
\nu_1 + \nu_2 + \nu \in\mathbb{Z}.
\end{equation}

Following~\cite{Speziale:2016axj}, the $\SL$-CG coefficients decompose into $\SU$- and boost-data according to
\begin{equation}
C^{\rho\nu jm}_{\rho_1\nu_1 j_1 m_1, \rho_2\nu_2 j_2 m_2}
=
\chi(\rho_1,\rho_2,\rho, \nu_1,\nu_2,\nu, j_1, j_2,j)C^{jm}_{j_1 m_1 j_2 m_2},
\end{equation}
where we write for brevity $\chi(\rho_1,\rho_2,\rho, \nu_1,\nu_2,\nu, j_1, j_2,j) \equiv \chi(j_1,j_2,j)$, which are complex functions specified in~\cite{Speziale:2016axj}. Then, integrals of the form
\begin{equation}
\int\limits_{\SL}\dd{h}\prod_i D^{(\rho_i,\nu_i)}_{j_i m_i l_i n_i}(h),
\end{equation}
which arise from the imposition of left $\SL$ invariance in Section~\ref{subsec:Spin Representation of the Action and Regularization}, as computed in~\cite{Speziale:2016axj}, yield
\begin{equation}\label{eq:Integral over four SL2C matrices}
\begin{aligned}
& \int\limits_{\SL}\dd{h}\prod_{i=1}^4 D^{(\rho_i,\nu_i)}_{j_i m_i l_i n_i}(h)\\[7pt]
& =
(-1)^{j_1-j_2+j_3-j_4-l_1+l_2-l_3+l_4}\sqrt{d_{j_4}}\sqrt{d_{l_4}}\int\dd{\rho_{12}}\sum_{\nu_{12}}4\left(\rho_{12}^2+\nu_{12}^2\right)\sum_{j_{12}l_{12}}\times\\[7pt]
& \times
\bar{\chi}(j_1, j_2, j_{12})\bar{\chi}(j_{12},j_3,j_4)\chi(l_1,l_2,l_{12})\chi(l_{12},l_3,l_4)\sqrt{d_{j_{12}}}\sqrt{d_{l_{12}}}\mqty(j_i \\ m_i)^{(j_{12})}\overline{\mqty(l_i \\ n_i)^{(l_{12})}},
\end{aligned}
\end{equation}
where $d_j$ denotes the dimension of the $j$-th $\SU$-representation, given by $(2j+1)$, $\mqty(j_i \\ m_i)^{(j_{12})}$ denotes $\SU$ intertwiner coefficients and $j_{12},~l_{12}$ are coupled $\SU$ spins, subject to $j_{12},l_{12}\geq\abs{\nu_{12}}$, resulting in the respective domains
\begin{align}
& j_{12} \in \{\max(\abs{\nu_{12}},\abs{j_1-j_2}),...,j_1+j_2\},\\[7pt]
& l_{12} \in \{\max(\abs{\nu_{12}},\abs{l_1-l_2}),...,l_1+l_2\}.
\end{align}
For notational ease and to suggest some resemblance with $\SU$-intertwiner coefficients, we define
\begin{equation}\label{eq:definition of SL2C intertwiner coefficients}
\begin{aligned}
& \sum_{j_{12}}(-1)^{-j_1 +j_2-j_3+j_4}\sqrt{d_{j_4}}\sqrt{d_{j_{12}}}\overline{\mqty(j_i \\ m_i)^{(j_{12})}}\chi(j_1,j_2,j_{12})\chi(j_{12},j_3,j_4)\\[7pt]
& \eqdef
\mqty((\rho_1,\nu_1) & (\rho_2,\nu_2) & (\rho_3,\nu_3) & (\rho_4,\nu_4) \\ j_1 m_1 & j_2 m_2 & j_3 m_3 & j_4 m_4)^{(\rho_{12},\nu_{12})}
\equiv
\mqty((\rho_i,\nu_i) \\ j_i m_i)^{(\rho_{12},\nu_{12})},
\end{aligned}
\end{equation}
for which Equation \eqref{eq:Integral over four SL2C matrices} can be simplified to
\begin{equation}\label{eq:simplified integral over four SL2C matrices}
\int\dd{h}\prod_i D^{(\rho_i,\nu_i)}_{j_i m_i l_i n_i}(h)
=
\int\dd{\rho_{12}}\sum_{\nu_{12}}4\left(\rho_{12}^2+\nu_{12}^2\right)\overline{\mqty((\rho_i,\nu_i) \\ j_i m_i)^{(\rho_{12},\nu_{12})}}\mqty((\rho_i,\nu_i) \\ l_i n_i)^{(\rho_{12},\nu_{12})}.
\end{equation}

We interpret the intermediate representation labels $(\rho_{12},\nu_{12})$ as $\SL$-intertwiner labels that will be associated to the coefficients of the condensate wavefunction after left-invariance is imposed in Section~\ref{subsec:Spin Representation of Condensate Wavefunction}.

For the regularization prescription in Section \ref{subsec:Spin Representation of the Action and Regularization} consider the following: Due to the non-compactness of the Lorentz group, the map
\begin{equation}
\int\limits_{\SL}\dd{h}\bigotimes_{i=1}^4\mathbf{D}^{(\rho_i,\nu_i)} \eqdef P : \bigotimes_{i=1}^4 \mathcal{D}^{(\rho_i,\nu_i)} \longrightarrow \text{Inv}_{\SL}\left(\bigotimes_{i=1}^4 \mathcal{D}^{(\rho_i,\nu_i)}\right),
\end{equation}
is not a projector, since $P^2 = P\cdot \text{vol}(\SL) \rightarrow\infty$, which will lead to a divergent kinetic term entering the action of the condensate wavefunction. However, we formally keep the diverging factor of vol$(\SL)$ and write $P$ as
\begin{equation}
P = \int\dd{\rho_{12}}\sum_{\nu_{12}}4\left(\rho_{12}^2+\nu_{12}^2\right)\ket{(\rho_{12},\nu_{12})}\bra{(\rho_{12},\nu_{12})},
\end{equation}
where $\ket{(\rho_{12},\nu_{12})}$ labels the basis vectors of the $\SL$-invariant subspace. Writing the $\SL$-intertwiner coefficients as
\begin{equation}
\mqty((\rho_i,\nu_i) \\ j_i m_i)^{(\rho_{12},\nu_{12})} = \braket{(\rho_i,\nu_i);j_i m_i}{(\rho_{12},\nu_{12})},
\end{equation}
we conclude that
\begin{equation}\label{eq:divergence of SL2C intertwiners}
\begin{aligned}
& \sum_{\{j_i m_i\}}\mqty((\rho_i,\nu_i) \\ j_i m_i)^{(\rho_{12},\nu_{12})}\overline{\mqty((\rho_i,\nu_i) \\ j_i m_i)^{(\rho_{12}',\nu_{12}')}}\\[7pt]
& =
\sum_{\{j_i m_i\}}\braket{(\rho_{12}',\nu_{12}')}{(\rho_i,\nu_i);j_i m_i}\braket{(\rho_i,\nu_i);j_i m_i}{(\rho_{12},\nu_{12})}\\[7pt]
& =
\braket{(\rho_{12},\nu_{12})}{(\rho_{12}',\nu_{12}')}\\[7pt]
& =
\frac{\delta(\rho_{12}-\rho_{12}')\delta_{\nu_{12},\nu_{12}'}}{4\left(\rho_{12}^2+\nu_{12}^2\right)}\cdot\text{vol}(\SL).
\end{aligned}
\end{equation}

From Barrett-Crane intertwiners defined in Eq.~\eqref{eq:definition of BC intertwiner} and $\SL$-intertwiners defined in Eq.~\eqref{eq:definition of SL2C intertwiner coefficients}, one can form the invariant symbols $\{10\rho\}_{\text{BC}}$ and $\{15\rho\}$ which enter the vertex term of the action of the condensate wavefunction,  defined as the contraction of five Barrett-Crane and $\SL$-intertwiners, respectively. Their explicit form is given by
\begin{equation}\label{eq:definition of BC 10p symbol}
\begin{aligned}
\{10\rho\}_{\text{BC}}
& \defeq
\left(\prod_{a=1}^{10}\sum_{l_a n_a}\right)(-1)^{\sum_a n_a+l_a}B^{\rho_1 \rho_2 \rho_3 \rho_4}_{l_1 n_1 l_2 n_2 l_3 n_3 l_4 n_4}B^{\rho_4 \rho_5 \rho_6 \rho_7}_{l_4 -n_4 l_5 n_5 l_6 n_6 l_7 n_7}\times\\[7pt]
& \times
B^{\rho_7 \rho_3 \rho_8 \rho_9}_{l_7 -n_7 l_3 -n_3 l_8 n_8 l_9 n_9}B^{\rho_9 \rho_6 \rho_2 \rho_{10}}_{l_9 -n_9 l_6 -n_6 l_2 -n_2 l_{10} n_{10}}B^{\rho_{10} \rho_8 \rho_5 \rho_1}_{l_{10} -n_{10} l_8 -n_8 l_5 -n_5 l_1 -n_1},
\end{aligned}
\end{equation}
or equivalently
\begin{equation}\label{eq:integral form of 10p}
\begin{aligned}
\{10\rho\}_{\text{BC}}
& =
\int\limits_{\SL^5}\left[\dd{h}\right]^5D^{(\rho_1,0)}_{0000}(h_5^{-1}h_1)D^{(\rho_2,0)}_{0000}(h_4^{-1}h_1)D^{(\rho_3,0)}_{0000}(h_3^{-1}h_1)\times\\[7pt]
& \times D^{(\rho_4,0)}_{0000}(h_2^{-1}h_1)D^{(\rho_5,0)}_{0000}(h_5^{-1}h_2)D^{(\rho_6,0)}_{0000}(h_4^{-1}h_2)D^{(\rho_7,0)}_{0000}(h_3^{-1}h_2)\times\\[7pt]
& \times
D^{(\rho_8,0)}_{0000}(h_5^{-1}h_3)D^{(\rho_9,0)}_{0000}(h_4^{-1}h_3)D^{(\rho_{10},0)}_{0000}(h_5^{-1}h_4),
\end{aligned}
\end{equation}
and 
\begin{equation}\label{eq:definition of SL2C 15p symbol}
\begin{aligned}
& \{15\rho\}\defeq
\left(\prod_{a=1}^{10}\sum_{j_a m_a}\right)(-1)^{\sum_a j_a+m_a}
\mqty((\rho_1,\nu_1) & (\rho_2,\nu_2) & (\rho_3,\nu_3) & (\rho_4,\nu_4) \\ j_1 m_1 & j_2 m_2 & j_3 m_3 & j_4 m_4)^{(\rho_{11},\nu_{11})}\times\\[7pt]
& \times
\mqty((\rho_4,\nu_4) & (\rho_5,\nu_5) & (\rho_6,\nu_6) & (\rho_7,\nu_7) \\ j_4 -m_4 & j_5 m_5 & j_6 m_6 & j_7 m_7)^{(\rho_{12},\nu_{12})}
\mqty((\rho_7,\nu_7) & (\rho_3,\nu_3) & (\rho_8,\nu_8) & (\rho_9,\nu_9) \\ j_7 -m_7 & j_3 -m_3 & j_8 m_8 & j_9 m_9)^{(\rho_{13},\nu_{13})}\times\\[7pt]
& \times
\mqty((\rho_9,\nu_9) & (\rho_6,\nu_6) & (\rho_2,\nu_2) & (\rho_{10},\nu_{10}) \\ j_9 -m_9 & j_6 -m_6 & j_2 -m_2 & j_{10} m_{10})^{(\rho_{14},\nu_{14})}
\mqty((\rho_{10},\nu_{10}) & (\rho_8,\nu_8) & (\rho_5,\nu_5) & (\rho_1,\nu_1) \\ j_{10} -m_{10} & j_8 -m_8 & j_5 -m_5 & j_1 -m_1)^{(\rho_{15},\nu_{15})},
\end{aligned}
\end{equation}
where the contraction entering the $\{15\rho\}$-symbol is similar to the contraction pattern of the $\{15j\}$-symbol from $\SU$ recoupling theory. However, it should be stressed that the $\{15\rho\}$-symbol does not satisfy all the symmetry properties of the $\{15j\}$ symbol. In particular, its symmetry properties strongly rely on the definition of the functions $\chi(j_1,j_2,j_3)$, as pointed out in~\cite{Speziale:2016axj}.

\section{Spin representation of the extended Barrett-Crane model}\label{appendix:Spin representation of the extended BC model}

\cite{Baratin:2011tx} provided an extended GFT formulation for the Riemannian BC spin foam model based on the compact group SO$(4)$. However, to appropriately incorporate the local gauge group of actual gravity, an extension to Lorentzian signature is required, which has not been accomplished up to now. In particular, a crucial step for the computations involving coherent condensate states is the spin representation of a right-covariant group field on the extended domain $\SL^4\times\HH$, which is achieved in this appendix for the first time.

Notice, that the covariance property \eqref{eq:covariance under right} implies the invariance of the four group elements under the diagonal right action of $\SU_X$, which will lead to an $\SU$-intertwiner in the final expansion. Furthermore, representation matrices of group elements are decomposed and coupled to matrices of the timelike normal in~\cite{Baratin:2011tx} by normal-dependent Clebsch-Gordan coefficients, stemming from a projector from $\text{SO}(4)$ representations to representations of the stabilizer subgroup. It is the task of the following appendix to find the analogue of the Clebsch-Gordan coefficients for the Lorentzian case. 

In Equation \eqref{eq:decomposition of SL2C reps} we have already shown that the $\SL$-representation space $\mathcal{D}^{(\rho,\nu)}$ decomposes into $\SU$-representation spaces $\mathcal{Q}_j$, with $j\in\{\abs{\nu},\abs{\nu}+1,...\}$. Instead, $\SL$-representation spaces can also be decomposed into representations of $\SU_{X}$, denoted by $\mathcal{Q}_j(X)$,
\begin{equation}
\mathcal{D}^{(\rho,\nu)}
\cong\bigoplus_{j=\abs{\nu}}^{\infty}\mathcal{Q}_j(X).
\end{equation} 
We denote by 
\begin{equation}
\ket{(\rho,\nu);X;jm}
\end{equation}
the basis vectors of $\mathcal{D}^{(\rho,\nu)}$, where the domain of $j$ is specified above and $m$ runs over the usual values $m\in\{-j,...,+j\}$. The additional label, given by the timelike normal $X$, specifies to which $\SU_X$ subspace the state refers. Vectors associated to spaces of different timelike normals $X,X'$ are related by the action of an $\SL$-Wigner matrix. For $X\in\HH$ there is an $\SL$-element $h$, such that $X' = h\cdot X$. In this case, we find~\cite{Livine:2002ak}
\begin{equation}\label{eq:relation of vectors of different normals}
\ket{(\rho,\nu);X';jm}
=
\mathbf{D}^{(\rho,\nu)}(h)\ket{(\rho,\nu);X;jm}.
\end{equation} 
It is important to note, that the representation matrices that label the space $\mathcal{Q}_j(X)$ are really $\SU_X$ indices, not to confuse with $\SU$-indices. Now, in the early work~\cite{Livine:2002ak} where projected spin networks are introduced, the projector
\begin{equation}
P_j^{(\rho,\nu)}(X): \mathcal{D}^{(\rho,\nu)}  \longrightarrow \mathcal{Q}_j(X),
\end{equation}
is explicitly given by 
\begin{equation}
P_j^{(\rho\,\nu)}(X) = (2j+1)\int\limits_{\SU_X}\dd{u}\bar{\chi}^j(u)\mathbf{D}^{(\rho,\nu)}(u),
\end{equation}
where $\mathbf{D}^{(\rho,\nu)}$ denotes an $\SL$-representation matrix. Computing the matrix elements of the projector serves as a quick check that it satisfies the correct properties
\begin{equation}\label{eq:matrix elements of projector}
\begin{aligned}
& \bra{(\rho,\nu);X;j'm'}P_j^{(\rho,\nu)}(X)\ket{(\rho,\nu);X;l'n'}\\[7pt]
& =
(2j+1)\sum_{m}\int\limits_{\SU_X}\dd{u}\overline{D^j_{mm}(u)}D^{(\rho,\nu)}_{j'm'l'n'}(u)\\[7pt]
& =
(2j+1)\sum_m \delta_{j',l'}\int\dd{u}\overline{D^j_{mm}(u)}D^{j'}_{m'n'}(u)
=
\delta_{j'l'}\delta_{j'j}\delta_{m'n'},
\end{aligned}
\end{equation}
where the last line is obtained from the orthogonality relation of $\SU_X$-Wigner matrices. It is clear that $P_j$ acts as the identity on subspaces $\mathcal{Q}_j(X)$ and annihilates any $\mathcal{Q}_{j'}(X)$ for $j\neq j'$. It is important to observe that two representation spaces $\mathcal{Q}_j(X),\mathcal{Q}_j(X')$ associated to different normals are not orthogonal subspaces of $\mathcal{D}^{(\rho,\nu)}$ and therefore, $P_j(X)$ does not annihilate the space $\mathcal{Q}_j(X')$.

Following the last line of Equation \eqref{eq:matrix elements of projector}, the projector is re-written to
\begin{equation}
P_j^{(\rho,\nu)}(X)
=
\sum_m \ket{(\rho,\nu);X;jm}\bra{(\rho,\nu);X;jm}.
\end{equation}
As a consequence, the desired equivalent of the Clebsch-Gordan coefficients appearing in~\cite{Baratin:2011tx} for a given normal $X = [a]$ is given by
\begin{equation}
\braket{(\rho,\nu);[e];ln}{(\rho,\nu);[a];kp}
=
\bra{(\rho,\nu);[e];ln}\mathbf{D}^{(\rho,\nu)}(a)\ket{(\rho,\nu);[e];kp}
=
D^{(\rho,\nu)}_{lnkp}(a),
\end{equation} 
where we have used relation \eqref{eq:relation of vectors of different normals}. The indices $(k,p)$ are coupled to the $4$-valent $\SU$-intertwiner, finally leading to the expansion of a right-covariant function on $\SL^4\times\HH$
\begin{equation}\label{eq:group field with right covariance}
\begin{aligned}
& \varphi(g_1, g_2, g_3, g_4;[a])\\[7pt]
& =
\left[\prod_{i=1}^4\int\dd{\rho_i}\sum_{\nu_i}4\left(\rho_i^2+\nu_i^2\right)\sum_{j_i m_i l_i n_i}\sum_{k_i p_i}\right]\sum_{j}\varphi^{\rho_i\nu_i,k_i j}_{j_i m_i}\left(\prod_i D^{(\rho_i,\nu_i)}_{j_i m_i l_i n_i}(g_i)D^{(\rho_i,\nu_i)}_{l_i n_i k_i p_i}(a)\right)\mqty(k_i \\ p_i)^{(j)},
\end{aligned}
\end{equation}
where we write for brevity
\begin{equation}
\varphi^{\rho_i\nu_i,k_i j}_{j_i m_i}
\equiv
\varphi^{\rho_1\nu_1\rho_2\nu_2\rho_3\nu_3\rho_4\nu_4,k_1 k_2 k_3 k_4 j}_{j_1 m_1 j_2 m_2 j_3 m_3 j_4 m_4}.
\end{equation}
One can show that this expansion satisfies all of the three desired properties, being right-covariance, invariance under diagonal right $\SU_X$ action and invariance under change of representative of the normal $[a] \rightarrow [au], u\in\SU_X$. While the first property is obvious from the direct coupling of $g_i$ and $a$, the latter two are a consequence of the $\SU$-intertwiner.

\bibliographystyle{JHEP}
\bibliography{references.bib}

\end{document}